\documentclass[a4paper,11pt]{article}
\usepackage{jinstpub} 
\usepackage{lineno}
\usepackage{tikz}
\usepackage{graphicx}
\usepackage{enumitem}
\usepackage{booktabs}

\title{\boldmath Development and performance test of p-type Silicon pad array detector}

\author[a]{Sawan}
\author[a]{G.~Tambave}
\author[b]{S.~Das}
\author[b]{A.~Chaudhry}
\author[b]{R.~Gupta}
\author[a]{V.~K.~S.~Kashyap}
\author[a,1]{B.~Mohanty\note{Corresponding author.}}
\author[a]{M.M.~Mondal}
\author[b]{S.~Mathur}
\author[b]{A.~Puri}
\author[a]{K.~P.~Sharma}
\author[b]{R.~Sharma}
\author[a]{R.~Singh}

\affiliation[a]{National Institute of Science Education and Research, An OCC of Homi Bhabha National Institute, Jatni, Odisha, India}
 \affiliation[b]{Semi-Conductor Laboratory, Ministry of Electronics
and Information Technology (MeitY), S.A.S.Nagar,
Mohali, Punjab 160071, India}

\emailAdd{bedanga@niser.ac.in}
     
\abstract{
This article reports on the development and comprehensive evaluation of p-type silicon detector arrays fabricated at the Semi-Conductor Laboratory (SCL), Mohali, India. The detectors consist of an 8~$\times$~9 array of 1~$\times$~1~cm$^2$ pads fabricated on 6-inch wafers and read out using the High Granularity Calorimeter Readout Chip (HGCROC). Electrical characterization of the detector through current vs. voltage (IV) and capacitance vs. voltage (CV) measurements demonstrated consistent breakdown and full depletion voltages across all pads, in agreement with Technology Computer-Aided Design (TCAD) device simulations. Laboratory measurements with a $^{90}$Sr source and beam tests at PS, CERN with 10 GeV pions, showed a clear Minimum Ionizing Particle (MIP) signal, well separated from the pedestal and uniform response of the pads with an average signal-to-noise (S/N) ratio above 5.5. The measured shower profiles with 2-4 GeV positron beams for various thicknesses of a tungsten absorber placed in front of the detector are found to be in agreement with the corresponding Geant4 simulations. The performance test results for the detector show that it is a promising candidate for the future ALICE upgrade detector named Forward Calorimeter (FoCal). The FoCal will have alternating layers of low and high-granularity silicon pad detectors with absorbers as a part of the electromagnetic segment, and along with its hadronic segment, will study the direct photons, neutral hadrons, vector mesons, and jets production in the low Bjorken-x region.
}
\arxivnumber{} 
\begin{document}
\maketitle
\flushbottom

\section{Introduction}

Silicon detectors are widely utilized across various fields, such as medical imaging, material science, nuclear physics, and particle physics~\cite{Si_detector_medicalImaging, Si_detector_application, Si_detector_highEnergyPhysics}. In high-energy physics, the importance of silicon detectors is particularly pronounced, as modern collider experiments operate in environments with intense radiation over extended periods with high particle flux. Silicon detectors are well-suited for such conditions due to their excellent radiation tolerance, as well as their high spatial and energy resolution~\cite{focaltdr}. Technologies such as silicon pixel and strip detectors have significantly improved spatial resolution, enabling high-precision particle tracking. Furthermore, recent advancements in silicon sensor technologies such as Monolithic Active Pixel Sensors (MAPS) and Silicon Photomultipliers (SiPMs) have improved overall detector performance by integrating signal amplification and processing directly within the silicon substrate. Silicon detectors are being actively used in sampling calorimeters, which consist of alternating layers of passive absorbers to produce electromagnetic showers and active silicon layers to detect the shower particles and measure their energy. This configuration allows to sample the shower as a function of absorber depth, enabling the reconstruction of shower profiles.~\cite{Clice_pads1, Clice_pads2, focal_ptype_published, pixel_paper1, Pixel_paper2_calice, sanjib_paper}. 

The ALICE Forward Calorimeter (FoCal) is one such detector being developed as a part of the ALICE upgrade for LHC Run 4 data-taking~\cite{focal_LOI, physics_of_focal, focaltdr}. Its primary objective is the study of low Bjorken-x physics, where gluon saturation effects predicted by quantum chromodynamics (QCD) remains unexplored. To access this kinematic region, the FoCal detector will be installed in the forward pseudorapidity range of 3.2 < $\mathrm{\eta}$ < 5.8 at a distance of 7~m from the ALICE interaction point, where it will enable measurements of direct photons, neutral hadrons, vector mesons, and jets. 

The FoCal consists of two components: a silicon-tungsten electromagnetic sampling calorimeter (FoCal-E) and a conventional hadronic calorimeter (FoCal-H). FoCal-E consists of two silicon detector technologies, one is low-granularity silicon pad detectors with an individual pad size of 1~$\times$~1 cm$^2$, used for the measurement of shower energy and position, and the other is high-granularity silicon pixel detectors (MAPS) with a single pixel size of about 30~$\times$~30 $\mathrm{\mu \mathrm{m}^2}$, used for precise particle tracking. The pad detectors are read~out using High Granularity Calorimeter Readout Chip (HGCROC)~\cite{HGCROCv2_paper} developed by the CMS collaboration, and the pixel detectors are read~out using ALICE Pixel Detector (ALPIDE)~\cite{ALPIDE_chip}. The full detector setup includes 18 layers of silicon pads and 2 layers of silicon pixels, each preceded by a tungsten (W) absorber of thickness $\sim$1X$_0$ placed in front of them.

This paper presents the design, fabrication, and performance evaluation of p-type silicon pad array detectors developed indigenously in India using 6-inch Si wafers at the Semi-Conductor Laboratory (SCL), Mohali. These detectors are read out using the HGCROC ASIC and are intended for application in the ALICE FoCal upgrade project. While similar p-type detectors have also been developed by Hamamatsu Photonics, Japan, for the same project~\cite{focaltdr, focal_ptype_published}, the present work represents a parallel and independent development effort undertaken in India. This includes localized fabrication, packaging, and characterization of the detectors, forming a critical component of India’s contribution to FoCal. Prior R\&D and beam tests focused on n-type pad array detectors of similar geometry~\cite{ntype_fabrication_paper, ntype_testbeam_niser}; however, a transition to p-type sensors has been adopted due to their superior radiation tolerance and compatibility with the final HGCROC version (HGCROCv3), which supports only p-type readout.

The subsequent sections in the paper provide a detailed discussion of the silicon pad array design and fabrication; laboratory characterization (capacitance-voltage and current-voltage measurements, plus performance tests using $\beta^-$ radiation from $^{90}$Sr; and performance tests at the Proton Synchrotron (PS), CERN using 10 GeV pion and 2-4 GeV positron beams.

\section{Detector design and fabrication}
\subsection{Si pad array design} \label{SubSec: Si design}
\begin{figure}[!htbp]
\centering
\includegraphics[width=0.85\textwidth]{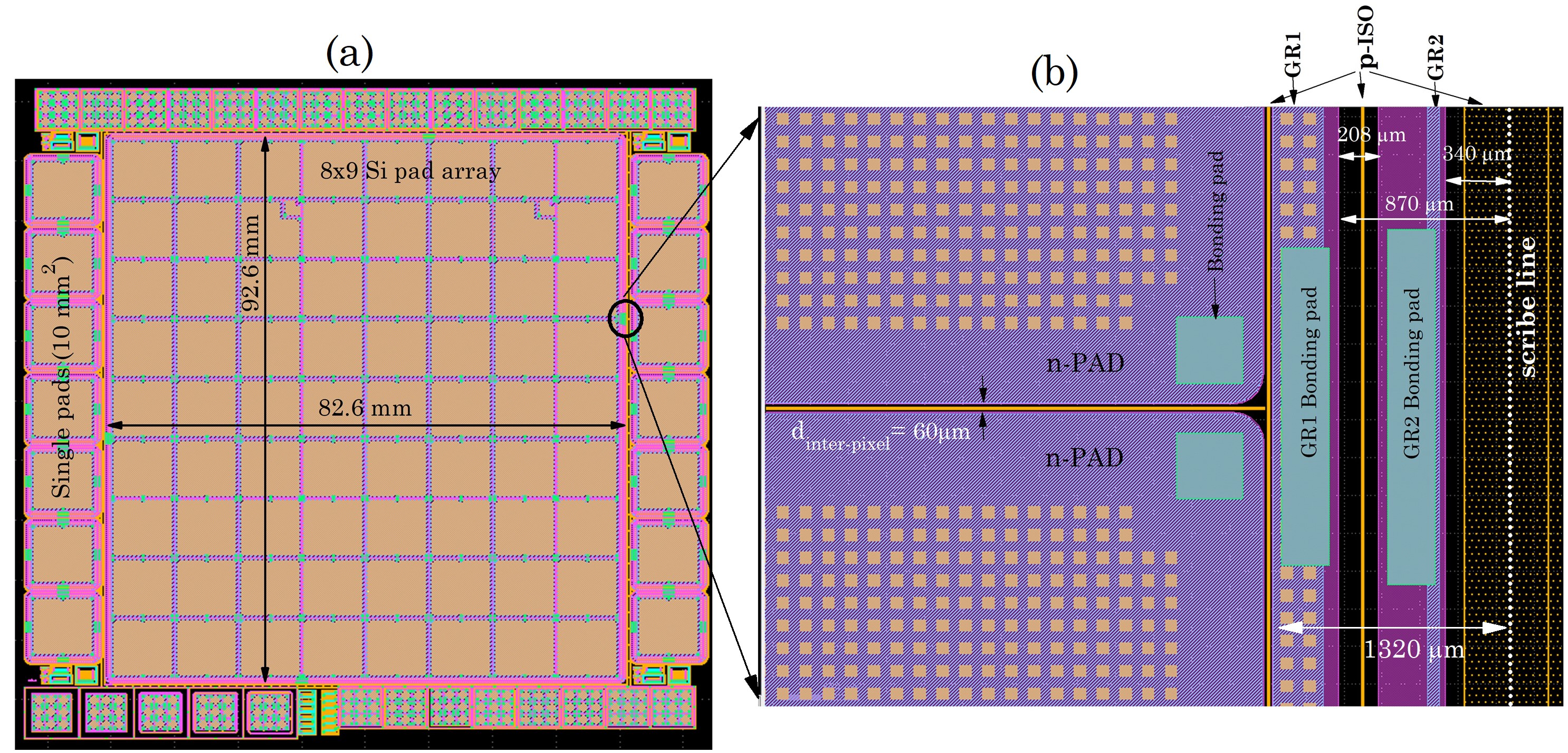}
\caption{(a) Layout showing 8$\times$9 pad array in the center and small pad cells in the periphery. Two calibration cells, each of size 3 $\times$ 3 $\mathrm{mm^2}$ are shown inside the pad cell. (b) Enlarged view of the outer edge of the pad array, showing top contact patterns (small brown squares), two guard rings, bonding pads for both the guard rings and the pad detectors, and their respective distances from the scribe line. \label{fig:Si_detector_layout}}
\end{figure}

The p-type silicon pad array detector is made from a 6-inch silicon wafer. It is segmented into a total of 72 pads in 8~$\times$~9 configuration, with each pad having a dimension of 1~$\times$~1 cm$^2$. Fig.~\ref{fig:Si_detector_layout}(a) shows the pad array design layout, with the main die ($82.6 \text{ mm} \times 92.6 \text{ mm}$) in the centre and single pads in the periphery, which are used for process characterization. Fig.~\ref{fig:Si_detector_layout}(b) shows the zoomed-in view around the outer edge of the pad array, where two guard rings are located, separated by 208~$\mu \mathrm{m}$. The pads are made by phosphorus implantation and are separated by a small distance of $60 \,\mu \mathrm{m}$. Additionally, the pads are isolated from each other using boron implantation, with each pad cell having six bonding points of size $350 \,\mu \mathrm{m}^2$.

\subsection{TCAD simulations} \label{SubSec: TCAD simulations}
\begin{figure}[!htbp]
\centering
\includegraphics[width=0.85\textwidth]{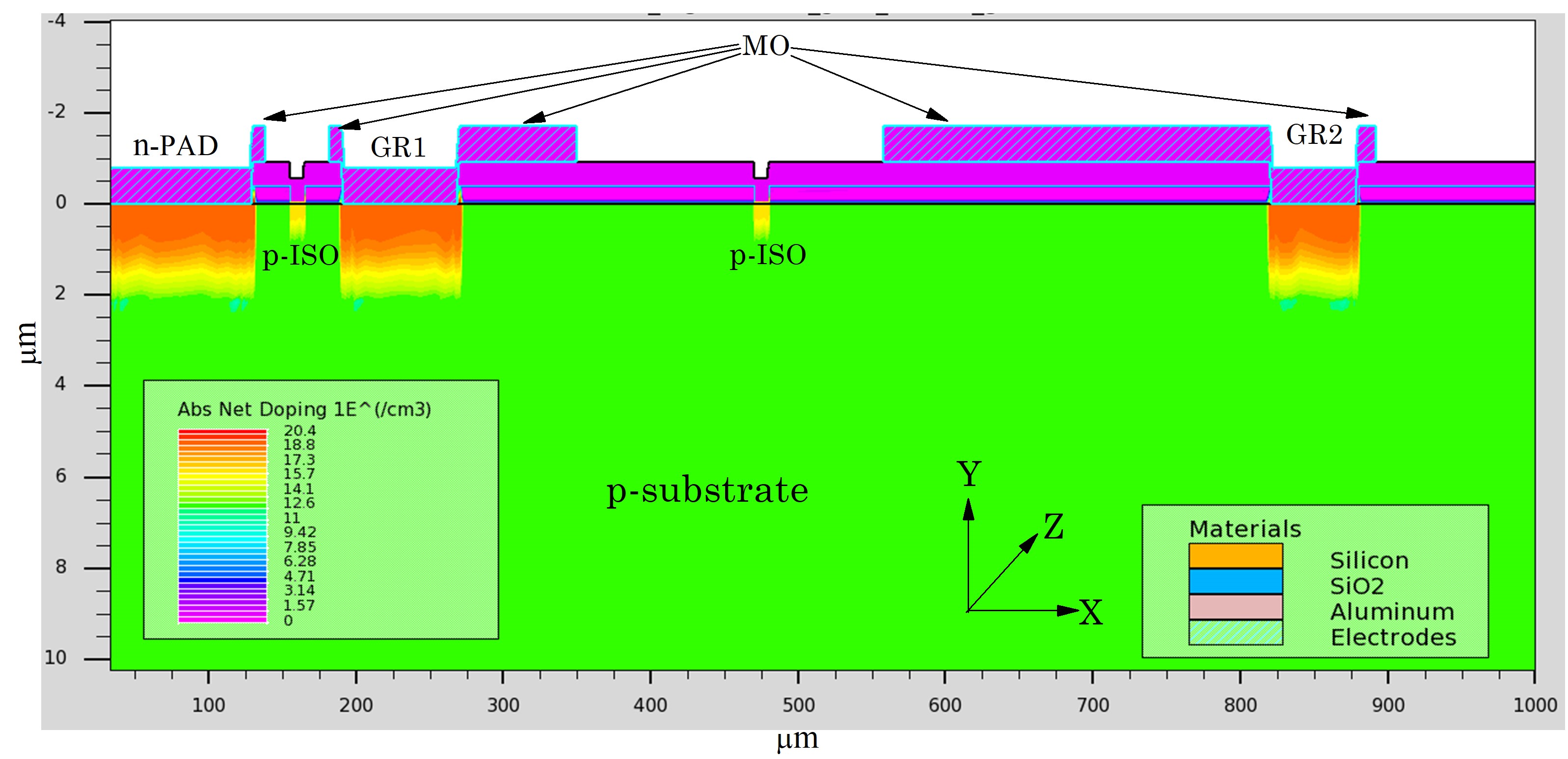}
\caption{The doping profile of the $8\times9$ Si pad array. This includes the part of the pad cell (n-PAD), two guard rings (GR1, GR2), and p-type isolation regions (p-ISO). The vertical axis represents depth in microns, while the horizontal axis shows the position in microns. The colour gradient indicates the absolute net doping concentration in cm$^{-3}$. MO denotes the metal overhang.
\label{fig:TCAD_simulations}}
\end{figure}

The physical characteristics of the detector are simulated using the Athena and Atlas modules of Silvaco Technology Computer-Aided Design (TCAD) software~\cite{TCAD_simulations}. The main design parameters are listed in the Tab.~\ref{tab:Process-parameters}. In the simulated doping profile shown in Fig.~\ref{fig:TCAD_simulations}, a high-resistivity p-type Silicon has been used as a substrate, resulting in $n^+/p^-/p^+$ vertical structure for the PADs and guard rings (p+ layer not shown). To reduce mesh size and computational time, only two pad cells at the edge, two guard ring structures, and the surrounding isolation regions were simulated. The colour legend on the left indicates the doping concentration. The pad cells and the guard rings are defined with the n-type phosphorus implantation, and isolation between them is formed by p-type boron implantation. The simulated junction depth exceeds 1~$\mu$m in both the pad cell and guard ring regions.

\begin{figure}[!htbp]
\centering
\includegraphics[width=0.45\textwidth]{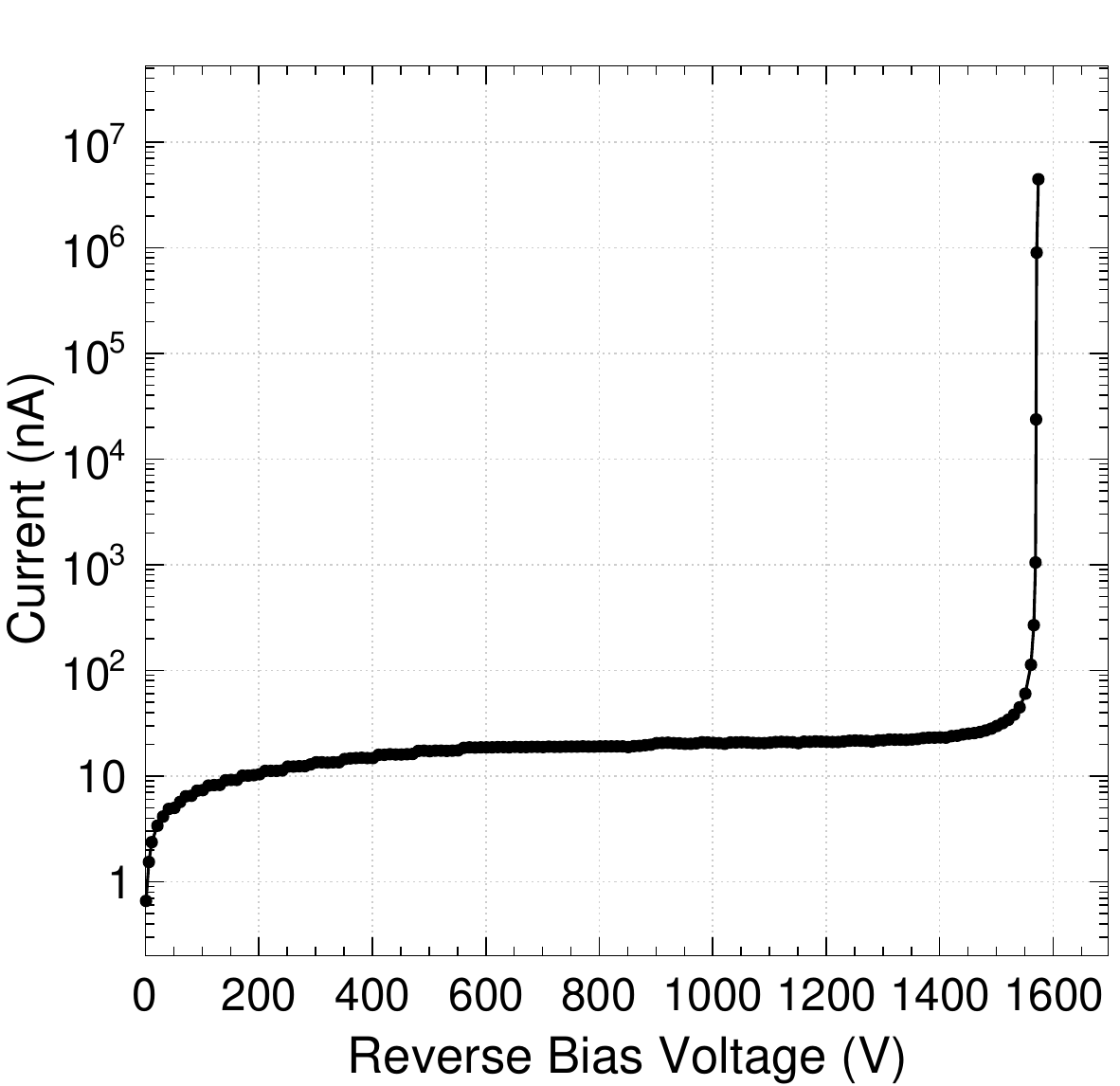}
\qquad
\includegraphics[width=0.45\textwidth]{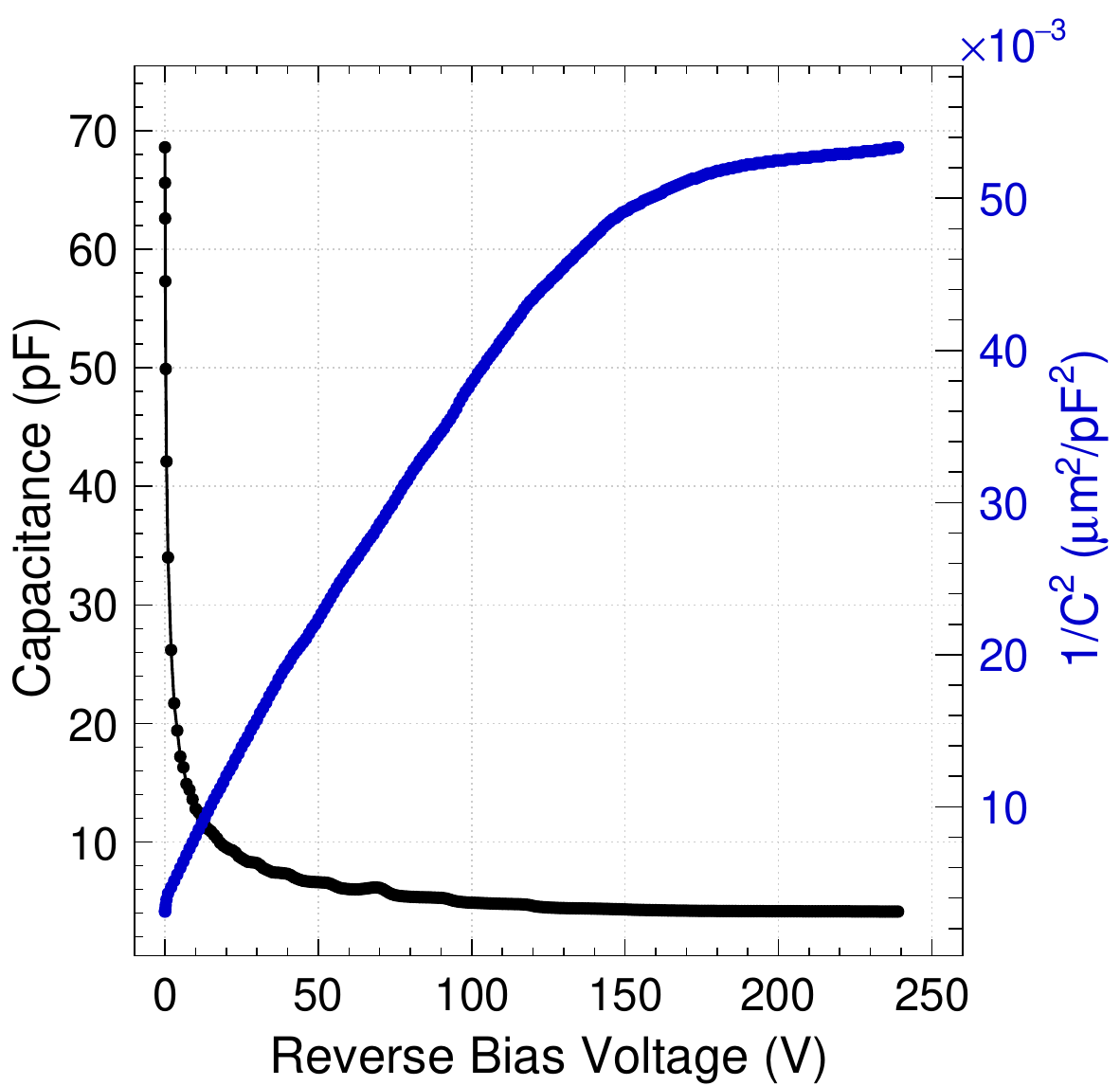}
\caption{(Left) Current–voltage (IV) and (right) capacitance–voltage (CV) characteristics of the single pad (1~$\times$~1 cm$^2$) obtained from TCAD simulations.
\label{fig:Sim_IV_and_CV}}
\end{figure}

The simulation also accounts for the effects of metal overhang on the pad detectors as well as on the guard rings. Various parameters such as implant dose, drive-in time and temperature, guard ring distance, metal overhang, etc., were optimised for the highest possible breakdown voltage and lowest leakage current. The pad cell and GR1 were kept at reverse bias with respect to the substrate to simulate the leakage current as well as the device capacitance as a function of reverse bias voltage. The simulated IV and CV plots are shown in Fig.~\ref{fig:Sim_IV_and_CV}. The IV characteristics indicate a breakdown voltage exceeding $1400 \,\text{V}$, while the CV characteristics suggest a full depletion voltage (FDV) between $160-170 \,\text{V}$.

\begin{table}[t]
\begin{center}
\caption{Design parameters of the Si pad array}
\begin{tabular}{||l||p{0.3\linewidth}||}
 \hline
 \textbf{Design/Process parameter} & \textbf{Values} \\  
 \hline\hline
 Die dimension & 82.6 mm $\times$ 92.6 mm \\ 
 \hline
 Single pad dimension & 9.94 mm $\times$ 9.94 mm \\
 \hline
 Pad pitch & 10 mm \\
 \hline
 Gap between pads & 60 $\muup$m \\
 \hline
  No. of Guard Rings (GR) &  2 N-type GR \\
 \hline
  Distance from scribe line to first pad & 1320 $\muup$m  \\
  \hline
  Breakdown voltage & 500-700 V \\
  \hline
  Full depletion voltage (FDV) & 160-170 V \\
  \hline
  Pad capacitance & $<$ 40~pF/cm$^2$ at FDV \\
  \hline
  Metal overhang & 8 $\muup$m \\
  \hline
  Radius of curvature of PAD corner & 75 $\muup$m \\
  \hline
  Field oxide thickness & 0.35 $\muup$m \\
  \hline
  P-N junction thickness & $\sim$ 1 $\muup$m \\
  \hline
  Front and back metal thickness & 1.0 $\muup$m \\
  \hline
  Passivation layer material & Low-Pressure Chemical Vapor Deposition (LPCVD) oxide \\
  \hline
  Passivation layer thickness & 3000 \r{A} \\
  \hline
  Max. misalignment between layers & $1-2 \,\mu \mathrm{m}$ \\
  \hline
\end{tabular}
\label{tab:Process-parameters}
\end{center}
\end{table}

\subsection{Si pad array fabrication}
\subsubsection{Fabrication in the SCL foundry} \label{SubSec: Fabrication}
Following the Si pad array design and the TCAD device simulations discussed in the previous sections, the p-type Si pad array detectors were fabricated on 6-inch p-type Silicon wafers. The fabrication process was carried out at the Semi-Conductor Laboratory (SCL), Mohali, Punjab, India~\cite{scl}. The details and specifications about the Si wafers and the photomasks are listed in Tab.~\ref{tab:si_properties} and Tab.~\ref{tab:mask_specifications}. A similar design and fabrication process for n-type Si pad array detectors on a 6-inch wafer is reported in~\cite{ntype_fabrication_paper}.

\begin{figure}[!htbp]
\centering
\includegraphics[width=0.85\textwidth]{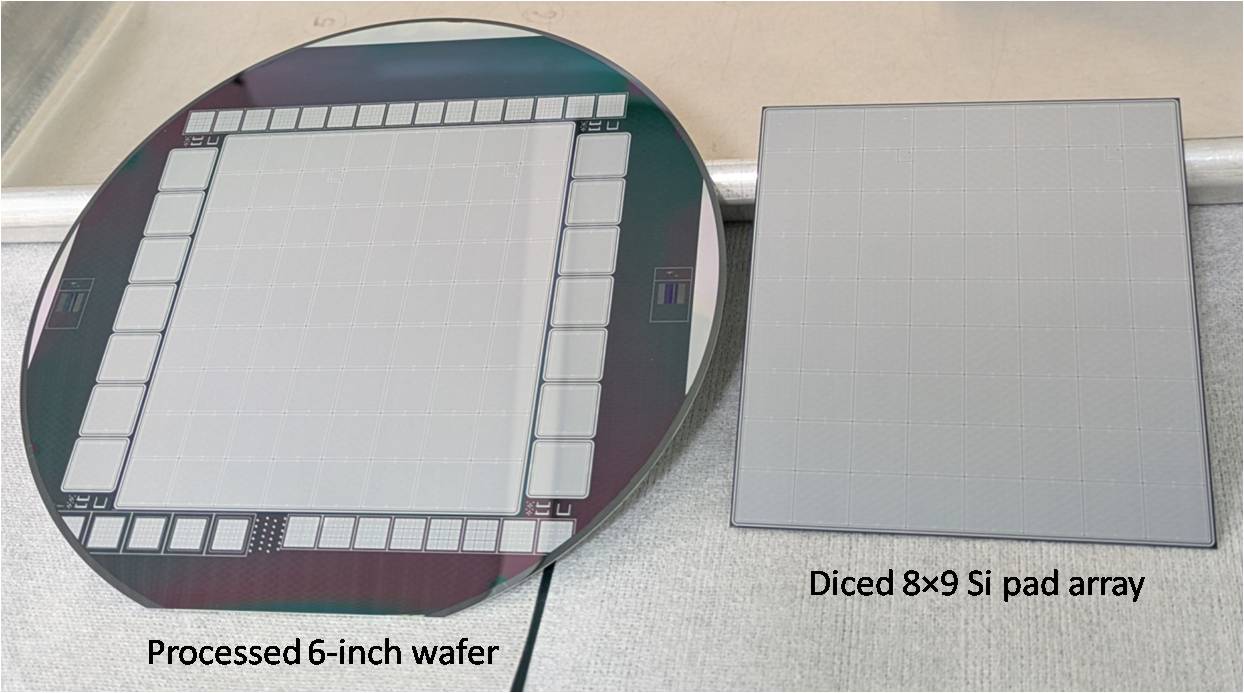}
\caption{Photographs show the full processed wafer (left) and diced 8$\times$9 Si pad array detector (right). 
\label{fig:wafer_photo}}
\end{figure}

The fabrication process involves six photomasks and a positive photoresist. All layers, except for the metal mask, are designed as dark field. Fabrication is carried out on a high-resistive thin Si wafer of $320 \,\mu \mathrm{m}$ thickness. The process flow is initiated with a sacrificial thermal oxidation to clean the incoming wafer surface. A field oxide layer, typically 3500~\r{A} thick, is then grown, followed by active area masking. Ion implantation with a positive photoresist as the mask is performed for the active area, guard rings, and inter-pad isolation. The active area and both the guard rings are formed by n-type implantation, while the inter-pad isolation and the isolation between the pad cells and guard rings are formed by p-type implantation. Dopant activation and deep junction formation are achieved through long-duration annealing cycles conducted in atmospheric thermal furnaces. A 3000 \r{A} Low-Pressure Chemical Vapor Deposition (LPCVD) oxide layer is then deposited for implant passivation. Contact openings for the device pads and guard rings are defined using a contact mask, followed by wet etching to form via holes. Subsequently, a $1 \,\mu \mathrm{m}$ thick aluminum layer is deposited and patterned using the metal mask, followed by dry etching. Metal overhang over oxide is ensured with the metal mask of about $8 \,\mu \mathrm{m}$ on either side of the pad cells. The aluminum layer is then passivated with a 3000 \r{A} Plasma-Enhanced Chemical Vapor Deposition (PECVD) oxide, which is finally patterned for etching pad contacts for wire bonding. The fabrication process is concluded with metal sintering at $440^\circ$C in an atmospheric thermal tube in a Nitrogen ambient. Each process step was optimized to minimize leakage current at full depletion and to maximize the breakdown voltage. A photograph of the finished wafer and a diced $8 \times 9$ Si pad array detector is shown in Figure~\ref{fig:wafer_photo}.

\begin{table}[htbp]
    \caption{Specification of Si wafers used for device fabrication.}
    \centering
    \begin{tabular}{|l|l|}
        \hline
        \textbf{Parameter} & \textbf{Values} \\
        \hline
        Type & p-type prime, Single crystal Si \\
        \hline
        Growth & Float Zone (FZ) \\
        \hline
        Diameter & $150 \pm 0.2$ mm \\
        \hline
        Orientation & $\langle 1 0 0 \rangle$ \\
        \hline
        Thickness & $320 \pm 15 \,\mu \mathrm{m}$ \\
        \hline
        Substrate resistivity & $\sim 5-9 \,\text{k}\Omega\text{-cm}$ \\
        \hline
        Dopant & Boron, p-type \\
        \hline
        Total Thickness Variation (TTV) & $<10 \,\mu \mathrm{m}$ \\
        \hline
        Maximum oxygen and carbon conc. & $<2 \times 10^{16} \text{/cm}^3$ \\
        \hline
        Minority carrier recombination lifetime & $>1$ ms \\
        \hline
    \end{tabular}
    \label{tab:si_properties}
\end{table}

\begin{table}[htbp]
    \centering
     \caption{Specification of photo-mask used during device fabrication.}
    \begin{tabular}{|l|l|}
        \hline
        \textbf{Parameter} & \textbf{Values} \\
        \hline
        Material & Soda lime glass / Anti-reflective chrome \\
        \hline
        Dimensions & $7'' \times 7'' \times 0.12''$ \\
        \hline
        Pattern generation & Direct write \\
        \hline
        Min. size of layout elements in layer & $10 \,\mu \mathrm{m}$ \\
        \hline
        Writing grids & $0.1 \,\mu \mathrm{m}$ \\
        \hline
        Mask order & 
        \begin{tabular}{@{}l@{}}
            (1) Active area mask \\
            (2) Active and guard ring implant mask \\
            (3) Isolation implant mask \\
            (4) Contact opening mask \\
            (5) Metal patterning mask \\
            (6) Pad opening mask
        \end{tabular} \\
        \hline
    \end{tabular}
    \label{tab:mask_specifications}
\end{table}

\subsubsection{Integration with readout electronics} \label{SubSec: Integration with HGCROC}
The p-type silicon pad arrays were interfaced with the highly integrated ASIC named High Granularity Calorimeter Readout Chip (HGCROC). The chip has 72 channels, with each channel consisting of a complete electronics chain, including a pre-amplifier, shaper, a 10-bit analog-to-digital converter (ADC) operating at a 40 MHz clock, and time-to-digital converters (TDCs) that allow time-over-threshold (TOT) measurements. The chip is capable of handling dynamic charge ranges from 0.2 fC to 10 pC, with noise levels below 0.4 fC~\cite{HGCROCv2_paper}. The HGCROC chip has a buffer memory of 512 samples~\cite{bourrion2023prototype}, which temporarily stores the data, and upon receiving a trigger signal, transmits the data to the KCU105 data acquisition (DAQ) board~\cite{KCU105} via high-speed 1.28 Gbps links for further processing and analysis. The HGCROC is mounted on a custom-designed printed circuit board (PCB), which is directly glued to the silicon pad arrays. More information about the readout electronics and their integration with the detectors is given in the reference~\cite{ntype_fabrication_paper}.

\section{Detector characterization using laboratory measurements} \label{Sec: lab measurements}
\subsection{IV and CV measurements} \label{Subsec: IV and CV data} 

\begin{figure}[!htbp]
\centering
\includegraphics[width=0.45\textwidth]{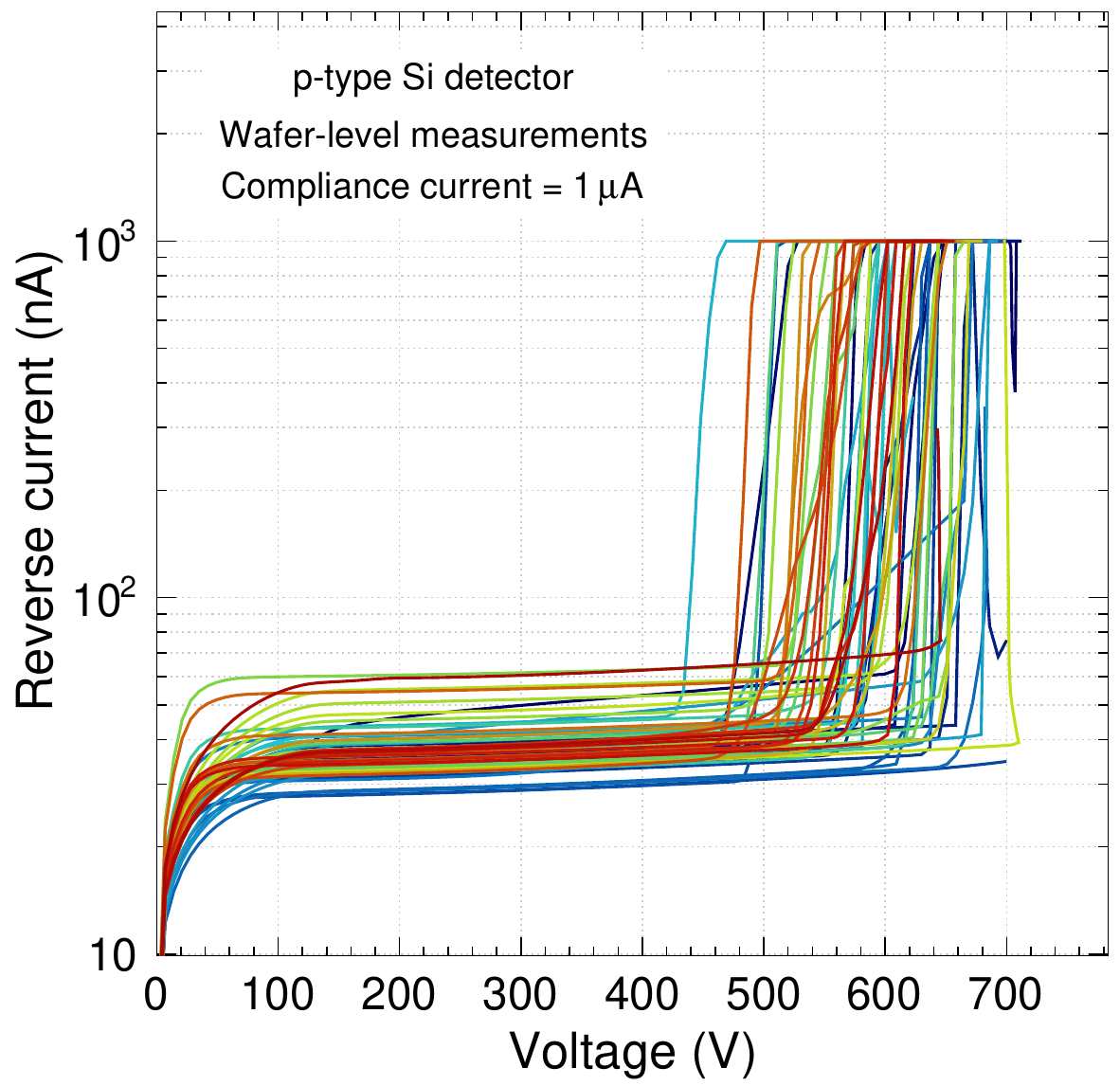}
\caption{Leakage current as a function of applied reverse bias voltage for various pad cells in the 8$\times$9 array. Measurements were performed on the silicon wafer prior to dicing and packaging with the compliance current set to 1$\mu$A.
\label{fig:wafer_level_IV}}
\end{figure}

The current vs. voltage (IV) and capacitance vs. voltage (CV) characterizations of the Si pad arrays were performed using a probe station and a Keysight B1505 high-voltage device analyzer before dicing. The pad under test was reverse biased from 0 to 700~V for the IV measurement with compliance current (maximum allowable current) set to 1~$\mu$A, and from 0 to 250~V for the CV measurement. All other pads and the inner guard ring (GR1) were biased together, synchronously swept with the same voltage as the test pad. The outer guard ring (GR2) was left floating throughout the measurements. The leakage current of various pad cells is plotted in Figure~\ref{fig:wafer_level_IV}. Around 92\% of the pad cells in the array exhibit leakage current below $50$ nA/cm$^2$. The breakdown voltage of the pads ranges from 450~V to around 700~V. The lower breakdown voltages observed in the data as compared to simulations are likely due to point defects or short-range extended defects~\cite{lower_breakdown2}.

\begin{figure}[!htbp]
\centering
\includegraphics[width=0.45\textwidth]{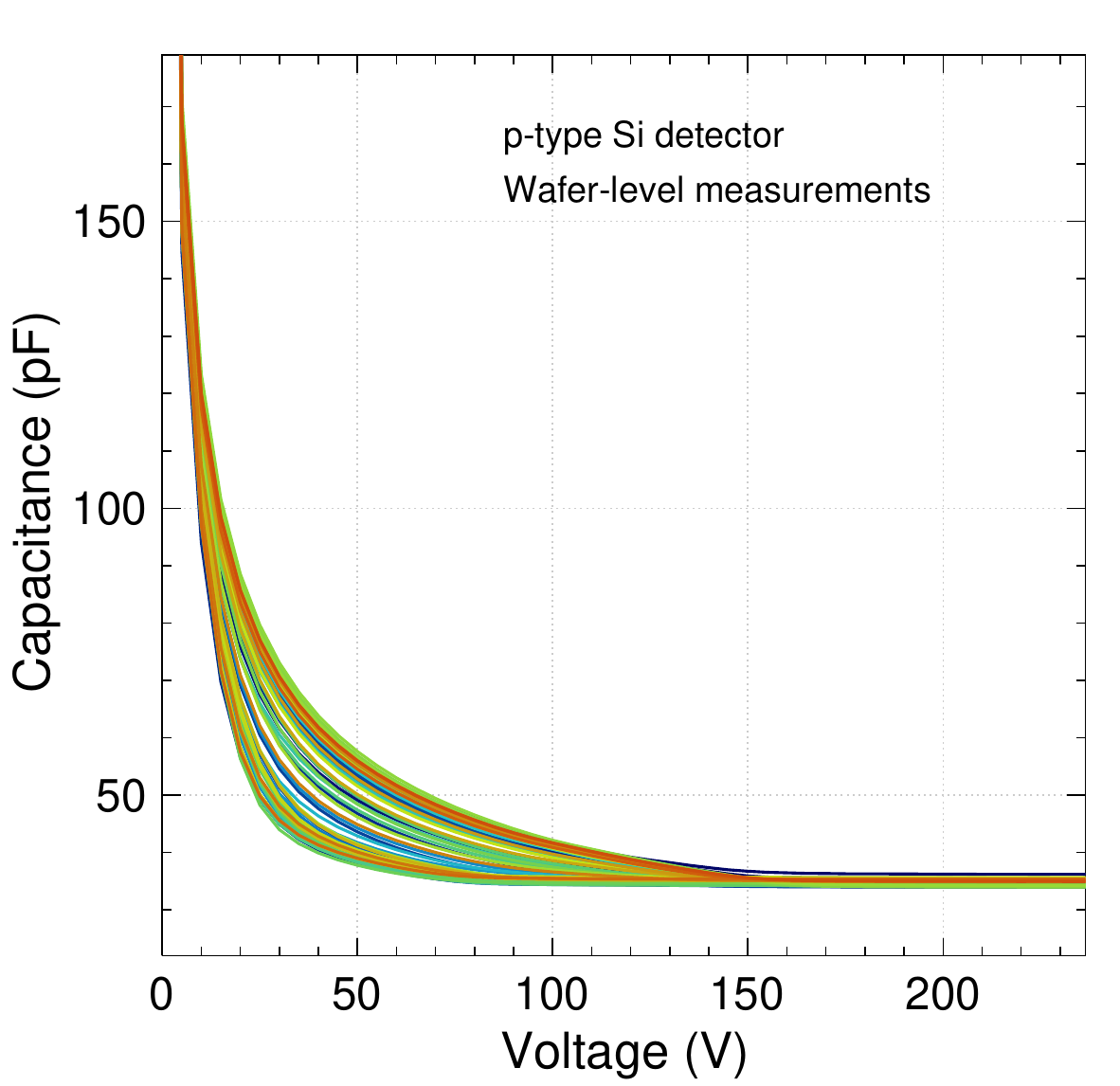}
\qquad
\includegraphics[width=0.45\textwidth]{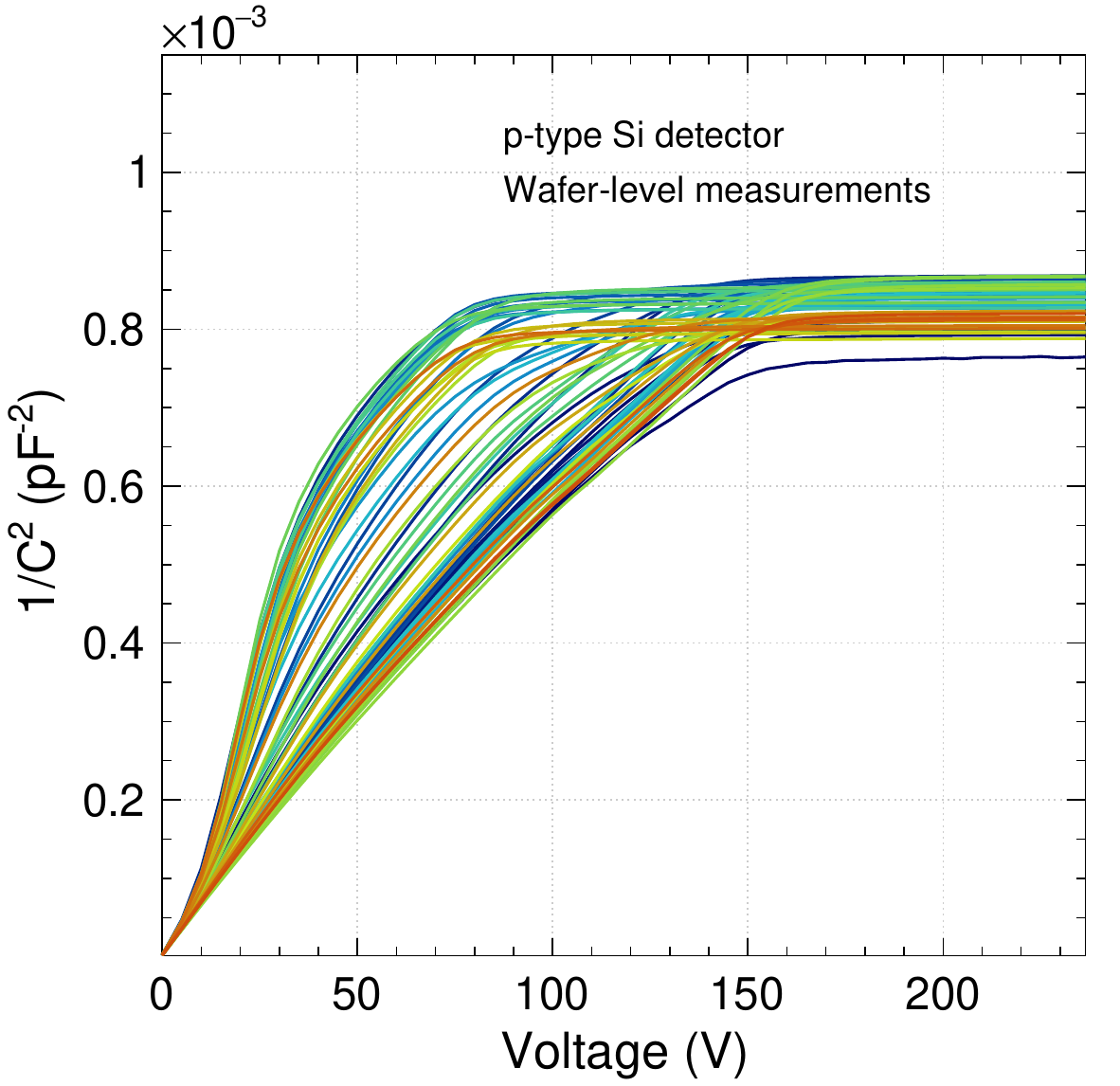}
\caption{(Left) Capacitance of the detector as a function of applied reverse bias voltage. (Right) Plot of $1/C^2$ versus reverse bias voltage, used to extract the full depletion voltage. Measurements were performed on the silicon wafer prior to dicing and packaging.
\label{fig:wafer_level_CV}}
\end{figure}

The CV characteristics for all pad detectors are shown in the left panel of Figure~\ref{fig:wafer_level_CV}. Above 150~V, the capacitance for all pads remains below 40~pF. The Full Depletion Voltage (FDV) corresponds to the voltage at which the $1/C^2$ curve saturates, which is found to lie within the range of 90--150~V (right panel of Fig.~\ref{fig:wafer_level_CV}), which is around one third of the observed breakdown voltage meeting the requirements for FoCal. The observed variation in FDV across pads may be attributed to non-uniform wafer resistivity. For stable operation during data acquisition, the detector is biased at 220~V, well above the full depletion voltage range.

\subsection{\texorpdfstring{Tests with $^{90}$Sr source}{Tests with 90Sr source}}

\begin{figure}[!htbp]
\centering
\includegraphics[width=0.425\textwidth]{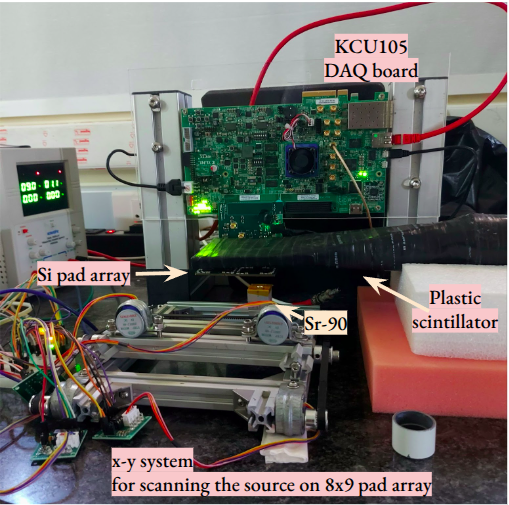}
\qquad
\includegraphics[width=0.45\textwidth]{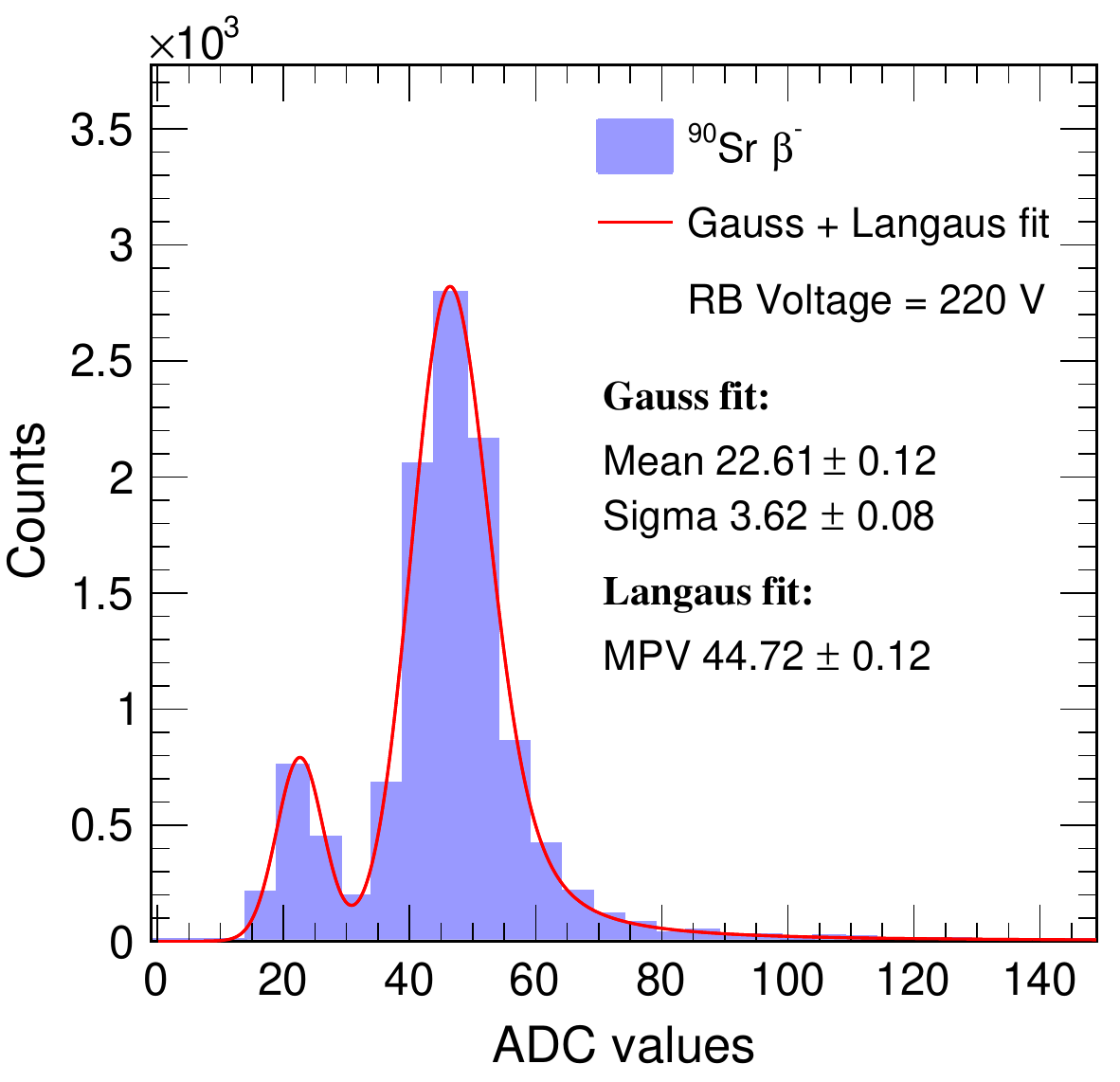}
\caption{(Left) Laboratory test setup for evaluating the detector performance using an $^{90}$Sr source enclosed in an aluminum box. The source is positioned below the detector on an XY scanner arrangement. The detector is connected to the DAQ board via an interface PCB. (Right) Response of single detector pad to $^{90}$Sr electron MIP.\label{fig:lab_test_setup_and_mip}}
\end{figure}

A $^{90}$Sr $\beta^{-}$ source which emits electrons of energy up to 2.28 MeV, was used to evaluate the performance of the packaged detector in the laboratory. A photograph of the test setup is shown in the left panel of Figure~\ref{fig:lab_test_setup_and_mip}. The $^{90}$Sr electron source is placed below the detector inside an aluminum box with a thickness of 6~mm, with a small aperture on the top to allow electrons to reach a single pad of the detector array. The detector was reverse-biased at 220~V using a Keithley SMU 2470. Data acquisition is done using a KCU105 DAQ board connected to the detector via an interface board. A plastic scintillator (PSc) provided the trigger signal to the DAQ board. The triggered data is then transmitted via Ethernet to a computer.

Using this setup, Minimum Ionising Particle (MIP) signals from $^{90}$Sr electrons were recorded for individual detector pads. An example distribution for one of the pads is shown in the right panel of Figure~\ref{fig:lab_test_setup_and_mip}. The distribution exhibits two distinct peaks: the left corresponds to the pedestal, while the right represents the electron MIP signal. It is fitted using a combination of a Gaussian function for the pedestal and a Langaus function (convoluted Landau and Gaussian functions) for the MIP signal. The Most Probable Value (MPV) extracted from the Langaus fit corresponds to the MIP peak, while the mean of the Gaussian fit represents the pedestal peak. A clear separation between the MIP peak and the pedestal is observed, with a signal-to-noise ratio of 6.1.

\subsubsection{Voltage scan} \label{subsec: Voltage scan lab}
\begin{figure}[!htbp]
\centering
\includegraphics[width=0.45\textwidth]{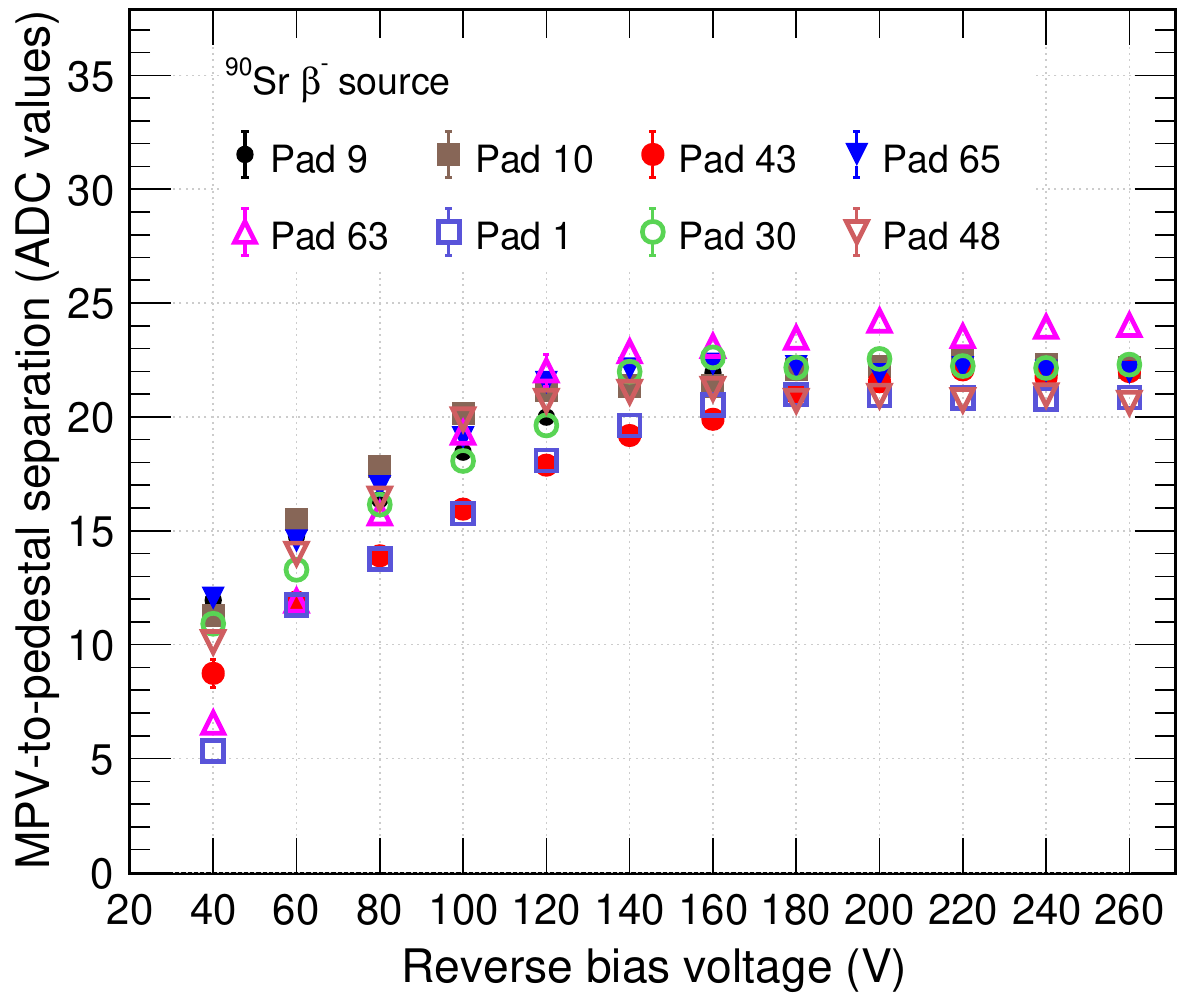}
\qquad
\includegraphics[width=0.4\textwidth]{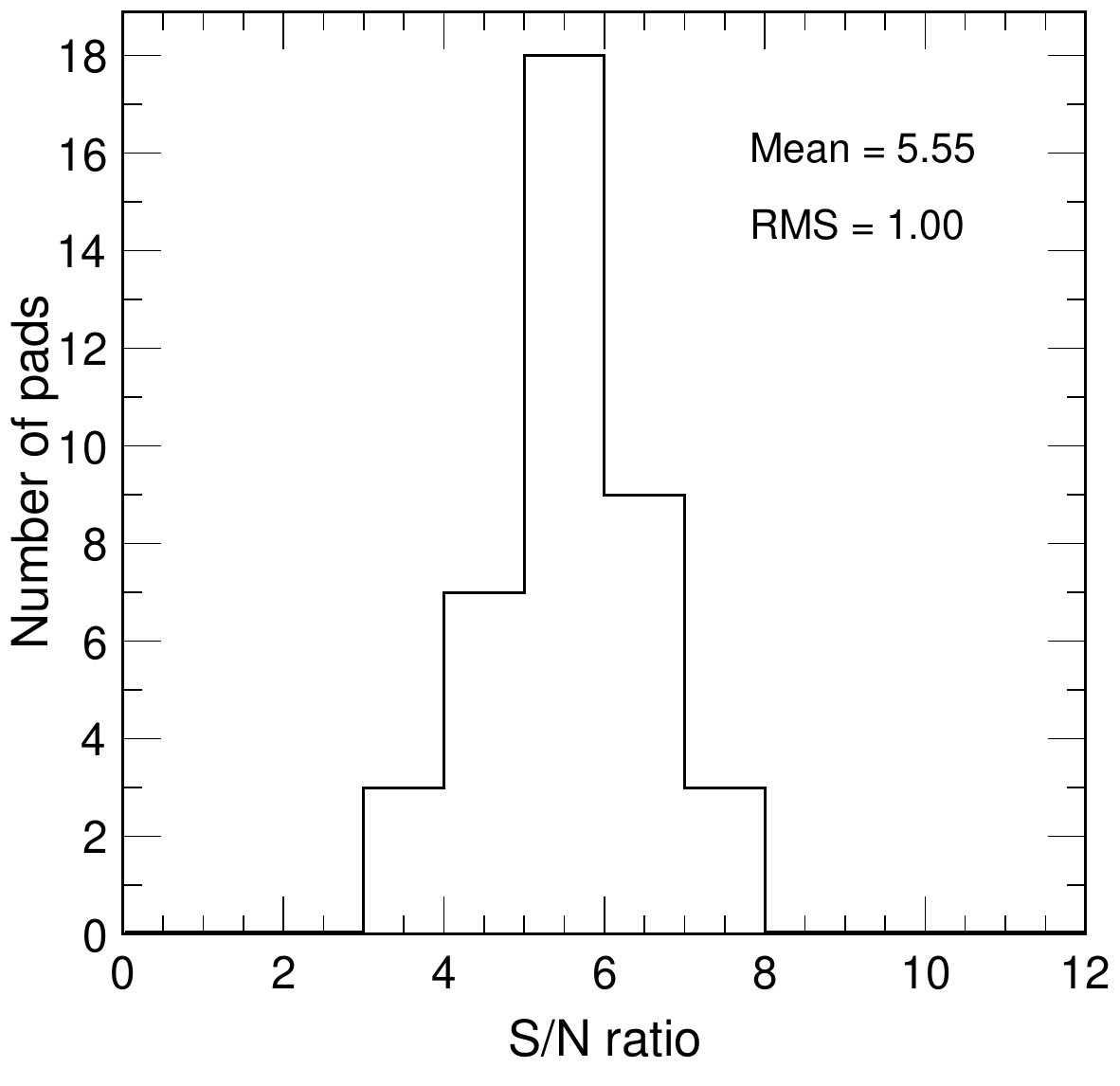}
\caption{(Left) $^{90}$Sr electron MPV-to-pedestal separation measured on a single detector pad as a function of applied reverse bias voltage (40–260~V). The error bars represent the quadrature sum of the uncertainties in the pedestal mean and MPV obtained from the fit. (Right) Signal-to-noise (S/N) ratio across different pads, measured during a position scan at a fixed reverse bias voltage of 220~V. \label{fig:voltage_and_position_scan_lab}}
\end{figure}

The MPV-to-pedestal separation is studied as a function of applied reverse bias voltage for eight pads from different regions over the detector array, as shown in the left panel of Fig.~\ref{fig:voltage_and_position_scan_lab}. The results demonstrate that the separation increases by increasing the reverse bias voltage up to 150~V and then starts to saturate. This behaviour is attributed to the widening of the depletion region with increasing reverse bias voltage, which enhances the collection efficiency of electron-hole pairs by reducing recombination losses. Once the detector is fully depleted, a further increase in bias voltage does not significantly affect the depletion width, leading to the observed saturation of MPV-to-pedestal separation. Based on the CV measurements and the voltage scan, a bias voltage of 220~V is selected for operating the detector such that the detector is fully depleted, and also the detector response is stable under voltage fluctuations without going into a breakdown region.

\subsubsection{Position scan}
To assess the homogeneity in the response of different detector pads, an XY positioning system was employed to scan different pads of the detector. The XY system consists of two stepper motors controlled by an ATmega 328p microcontroller (Arduino Uno). The $^{90}$Sr source was placed inside an aluminum box with a 5~mm opening (collimator), which was mounted on top of the XY system. The whole XY arrangement was then positioned below the detector for scanning of different detector pads. The MIP signals from 40 pads were recorded, and their signal-to-noise (S/N) ratios are plotted in the right panel of Fig.~\ref{fig:voltage_and_position_scan_lab}, showing an average S/N ratio of 5.5. These results demonstrate the detector’s capability to resolve weak signals. In addition to the 40 pads shown, 7 HGCROC channels were non-functional, and 4 pads exhibited no response after the detector packaging with the chip. The remaining 21 pads showed MIP signals merged with the pedestal. The number of pads with acceptable S/N ratios can potentially be increased by operating the detector at lower temperatures, which helps to reduce the leakage current, as discussed in the next section. Given that this was the first iteration of the detector fabrication, further TCAD simulations and process optimizations are underway based on the current findings. These include implementing three guard rings instead of the current two, with the goal of improving both the S/N performance and the overall yield of functional pads in the array.

\subsubsection{Temperature scan}
\begin{figure}[!htbp]
\centering
\includegraphics[width=0.45\textwidth]{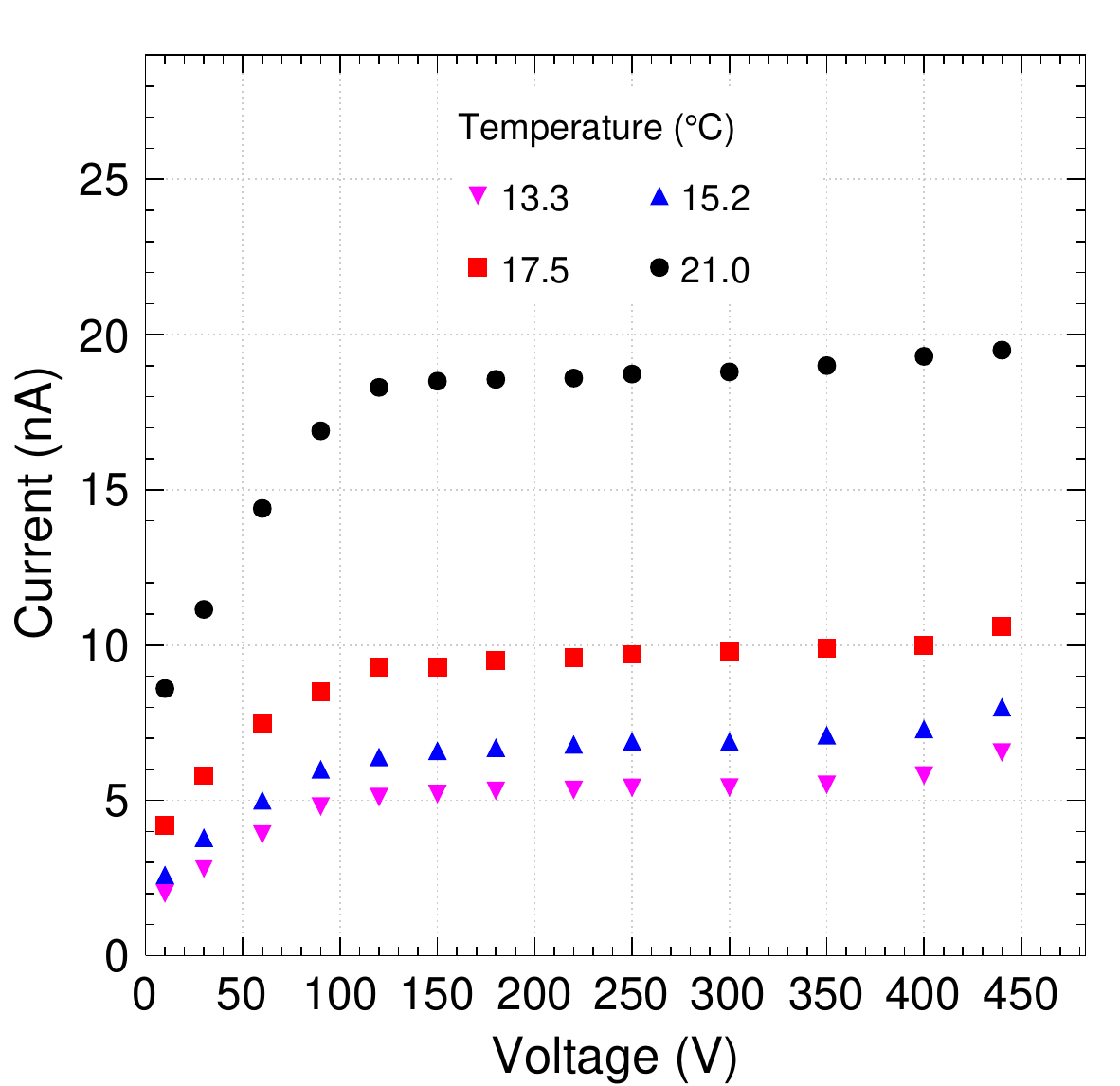}
\caption{Leakage current as a function of operating temperature for the detector.\label{fig:temperatuere_dependence}}
\end{figure}

Optimizing the leakage current is critical for semiconductor detectors operated in reverse bias mode, as thermally generated electron-hole pairs significantly contribute to the leakage current. To study the temperature dependence of the Si pad array detector, a Peltier module was used to control the detector temperature in the range of 13–25$^\circ$C. Fig.~\ref{fig:temperatuere_dependence} shows the variation of leakage current with temperature for bias voltages between 10 and 400~V. As the temperature is lowered from 21.0$^\circ$C to 13.3$^\circ$C, the leakage current decreases by approximately 12~nA at FDV. The FoCal detector is expected to operate at around 15$^\circ$C. Given that the HGCROC can tolerate leakage currents up to 50~$\mu$A per channel, the detector pads exhibiting leakage currents below 10~nA at the expected operating temperature are well within the chip requirements.

\section{Detector performance tests at PS CERN} \label{Sec: PS tests}
After thorough testing in the laboratory with $^{90}$Sr $\beta^{-}$ source, the detector was taken to the T10 facility at Proton Synchrotron (PS), CERN in Geneva, Switzerland, where it was tested with 10 GeV pion and 2-4 GeV positron beams.

\subsection{PS T10 beam characteristics}
The T10 beamline delivers secondary charged particles ($\mathrm{\mu , \,  \pi , \, p , \, e}$) with momenta ranging from 0.5 GeV/c to 7 GeV/c.  These secondary particles are generated by the interaction of a 24~GeV/c primary proton beam with a multi-target system~\cite{T10_beam}. The beam intensity ranges from 10$^3$ to 10$^6$ particles per spill, with a spill duration of approximately 400~ms and a repetition period of 20~s. The multi-target system has five different configurations~\cite{mixed-target}. For electron production, a cylindrical target configuration is used, consisting of 200~mm of Beryllium (Be) and 3~mm of Tungsten (W), both with a 10~mm diameter. For pion production, a hadron target, consisting of 200~mm of aluminum with a 10~mm diameter, is used. To further increase the purity of the beam, a Cherenkov detector is used.

\subsection{Experimental test beam setup}

\begin{figure}[!htbp]
\centering
\includegraphics[width=0.62\textwidth]{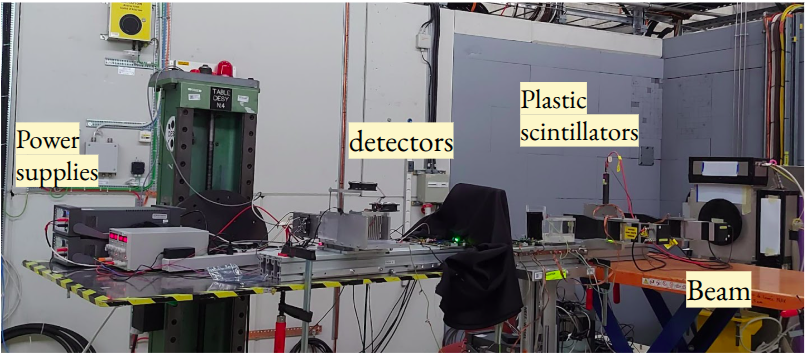}
\includegraphics[width=0.36\textwidth]{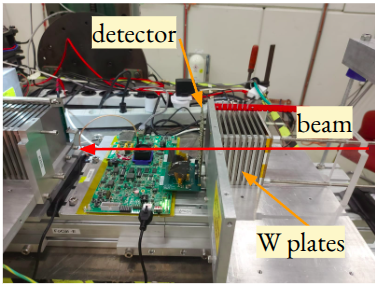}
\caption{(Left) A photograph of the test setup at T10 beamline in PS, CERN. The beam direction is from right to left. Four plastic scintillators, along with the Cherenkov detector, are used for the trigger logic. The detector is reverse-biased using the constant voltage supply from the source meter. (Right) Zoomed in view of the mechanical structure holding the detector and tungsten absorber plates.\label{fig:test_beam_setup}}
\end{figure}

Figure~\ref{fig:test_beam_setup} shows the experimental test beam setup at the PS, CERN. The detector was mounted on an adjustable table, enabling precise alignment with the beam trajectory in both the vertical and horizontal directions. An aluminum structure holding tungsten plates is kept in front of the detector to obtain the electromagnetic showers. The entire detector setup is covered with a black cloth to reduce leakage current caused by nearby lights. Signals generated from the combination of four plastic scintillators and a Cherenkov detector were fed to a discriminator configured with AND logic, followed by a NIM-to-TTL converter. The resulting TTL signal was used to trigger the DAQ board. Of the two larger plastic scintillators, one was fixed to the table, while the other was placed on a movable table near the detector. Two smaller "finger" scintillators, with a 1 cm$^2$ overlap area, ensured that the beam was incident on a single 1 cm$^2$ detector pad. This trigger configuration effectively suppressed background events, such as cosmic muons or particles from nearby beamlines. A zoomed-in view of the setup is shown in the right panel of Figure~\ref{fig:test_beam_setup}.

\subsection{Geant4 simulations}
\begin{figure}[!htbp]
\centering
\includegraphics[width=0.7\textwidth]{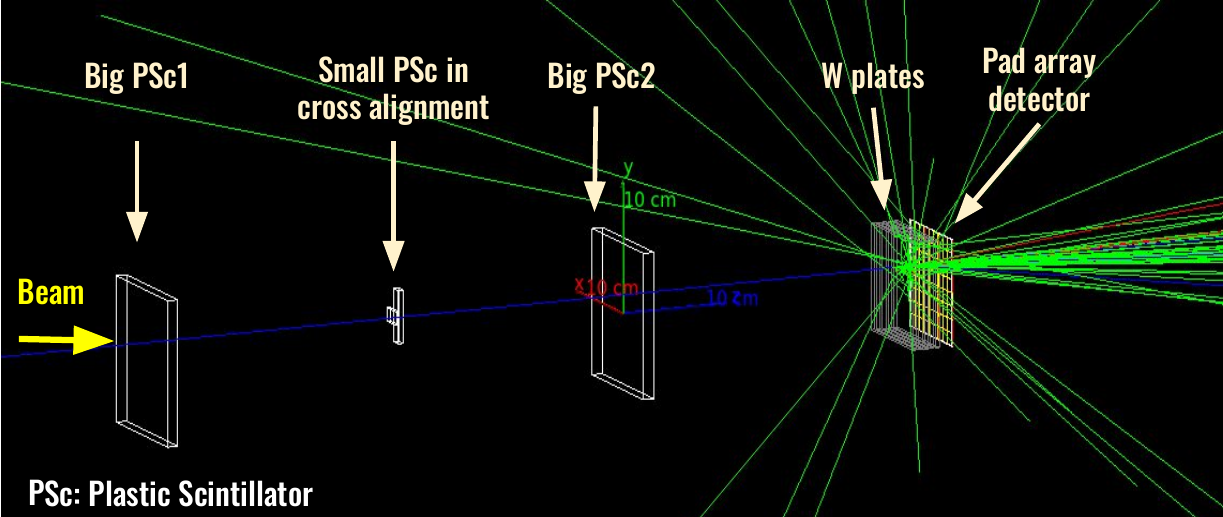}
\caption{Geant4 simulation of the test beam setup. The beam direction is from left to right. Detector segmentation (9$\times$8) is visualized by the hollow representation of the material. The coordinate system origin is defined at the upstream face of the larger plastic scintillator (PSc2), with the X, Y, and Z axes indicated by red, green, and blue arrows, respectively.\label{fig:geant4_sim_setup}}
\end{figure}

The test beam detector setup is implemented in Geant4 simulations~\cite{geant4package}, as shown in Figure~\ref{fig:geant4_sim_setup}. A 3~GeV e$^+$ beam (blue trajectory), depicted for a single event, propagates from left to right along the z-axis. The beam initiates an electromagnetic shower upon interacting with the absorbers, generating secondary particles. The particles are colour-coded in the Geant4 simulation: green for neutral particles, red for positively charged particles, and blue for negatively charged particles. The geometry and distances of the setup are summarized in Tab.~\ref{tab:dimensions_test_beam}.
\begin{table}[!htbp]
    \centering
        \caption{Dimensions (x~$\times$~y $\times$ z) of the components in the test beam setup. $\textit{PSc}$ refers to plastic scintillators. In the simulations, the beam propagates along the z-axis from left to right.}
    \renewcommand{\arraystretch}{1.2} 
    \begin{tabular}{||l||p{0.3\linewidth}||}
        \toprule
        \textbf{Component} & \textbf{Dimensions (mm)} \\
        \midrule
        Big PSc & 110 × 100 × 10 \\
        Finger horizontal PSc & 20 × 10 × 3 \\
        Finger vertical PSc & 50 × 10 × 5 \\  
        Tungsten absorber plates & 100 × 100 × 10 \\
        Si detector & 100 × 100 × 0.325 \\
        PSc 2 (origin) to PSc 1 separation & 460 \\
        PSc 2 (origin) to small PSc & 220 \\
        PSc 2 (origin) to Si detector & 720 \\
        \bottomrule
    \end{tabular}
    \label{tab:dimensions_test_beam}
\end{table}

\subsection{Detector response to pion beam}
\subsubsection{Voltage scan}
\begin{figure}[!htbp]
\centering
\includegraphics[width=0.4\textwidth]{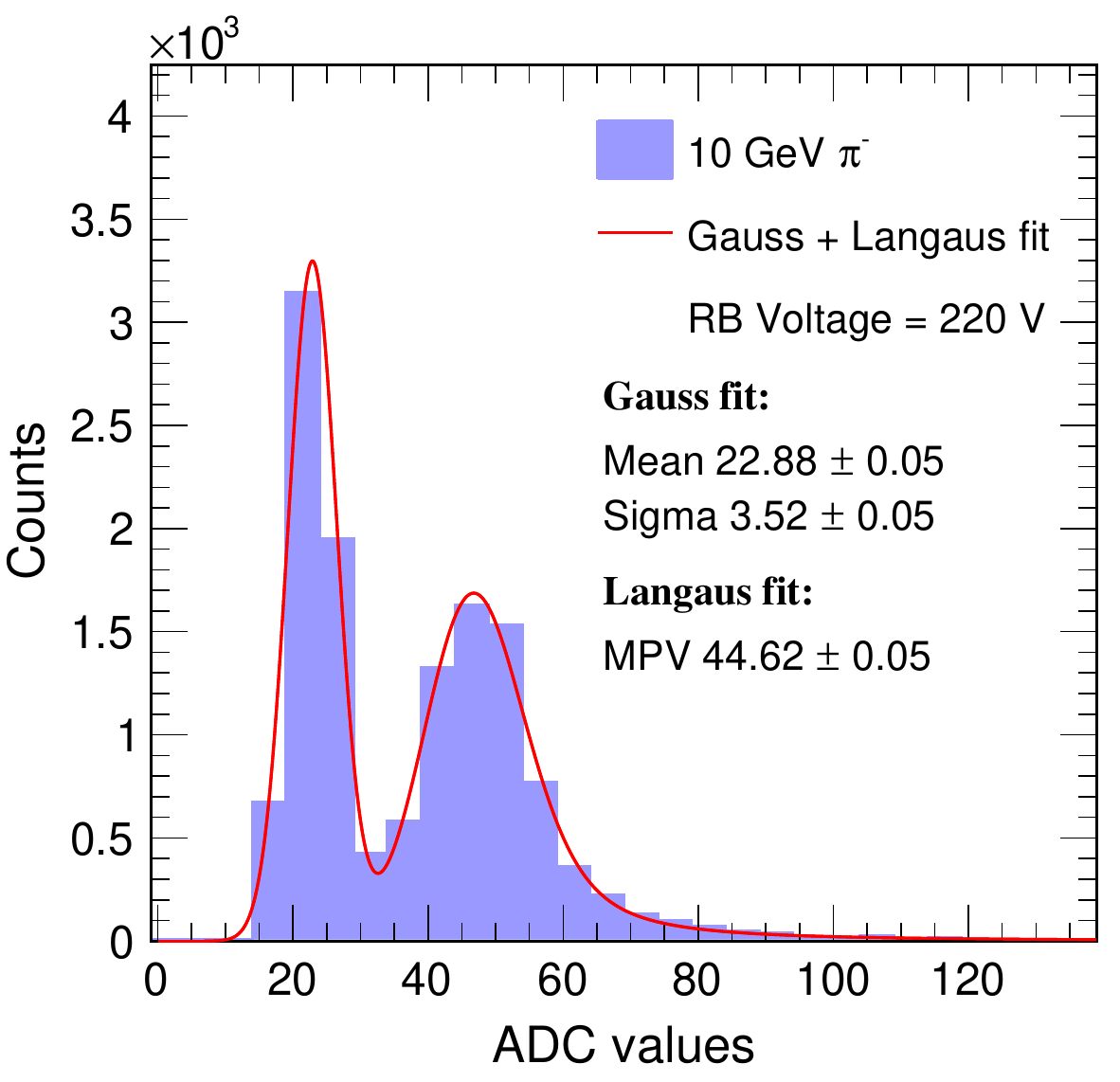}
\qquad
\includegraphics[width=0.43\textwidth]{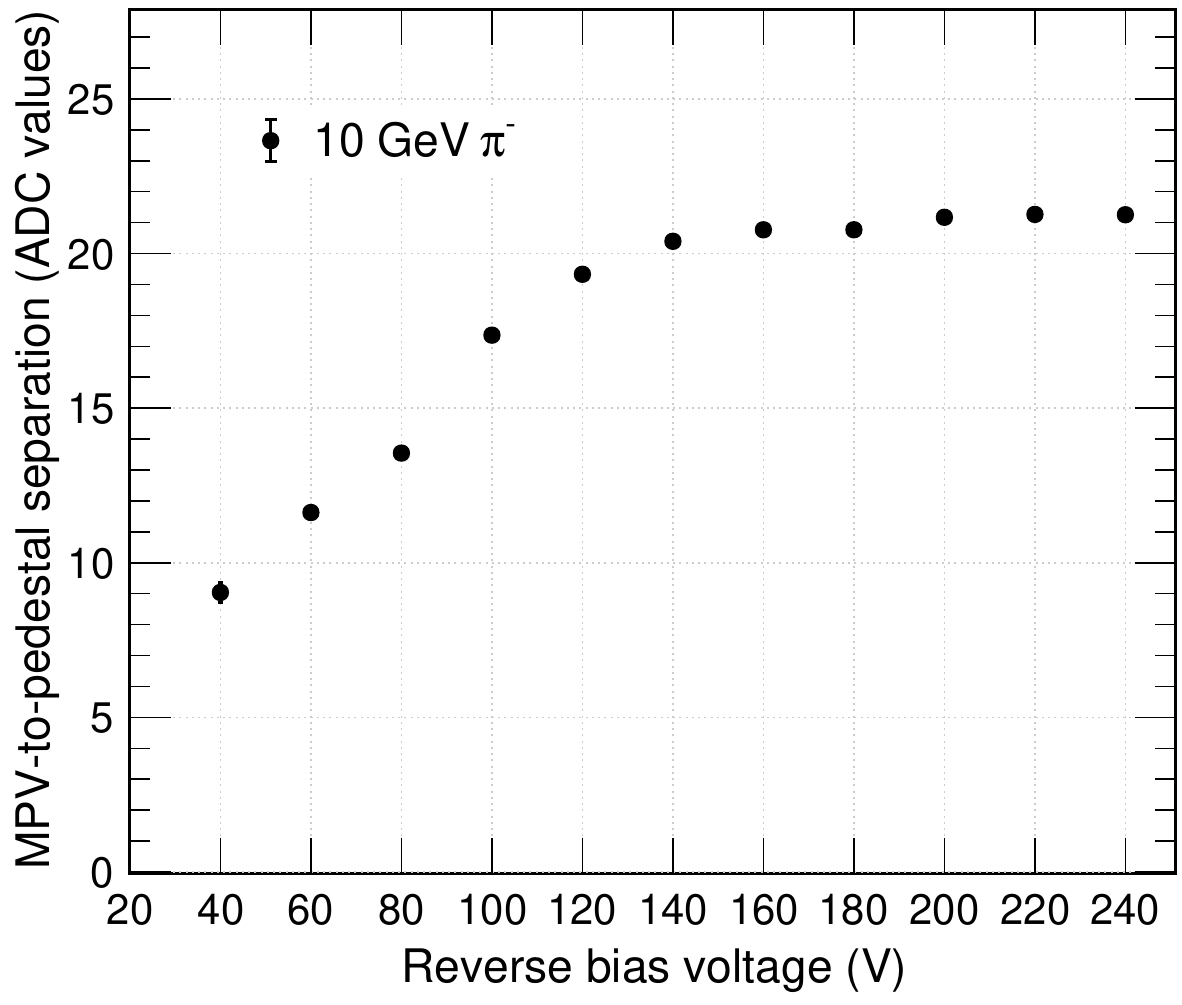}
\caption{(Left) Pion MIP spectrum at 220~V reverse bias.  The error bars represent the quadrature sum of the uncertainties in the pedestal mean and MPV obtained from the fit. (Right) Detector response to pion MPV-to-pedestal separation as a function of reverse bias voltage varied from 40~V to 240~V.\label{fig:voltage_scan_beam}}
\end{figure}

To characterize the response of the detector to minimum ionizing pions, measurements were performed using a 10~GeV pion beam. The resulting MIP spectrum at a reverse bias of 220~V for a single detector pad is shown in the left panel of Fig.~\ref{fig:voltage_scan_beam}. A clear separation between the MIP and pedestal peaks is observed, consistent with the results from the $^{90}$Sr source tests. The dependence of MPV-to-pedestal separation on reverse bias voltage was examined by varying the reverse bias from 40~V to 240~V, as shown in the right panel of Fig.~\ref{fig:voltage_scan_beam}. The results show that the separation increases with bias voltage and saturates around 150~V, indicating that the detector reaches full depletion around this voltage, consistent with the CV characterization (Sec.~\ref{Subsec: IV and CV data}) and laboratory measurements (Sec.~\ref{subsec: Voltage scan lab}).

\subsubsection{Position scan}
\begin{figure}[!htbp]
\centering
\includegraphics[width=0.445\textwidth]{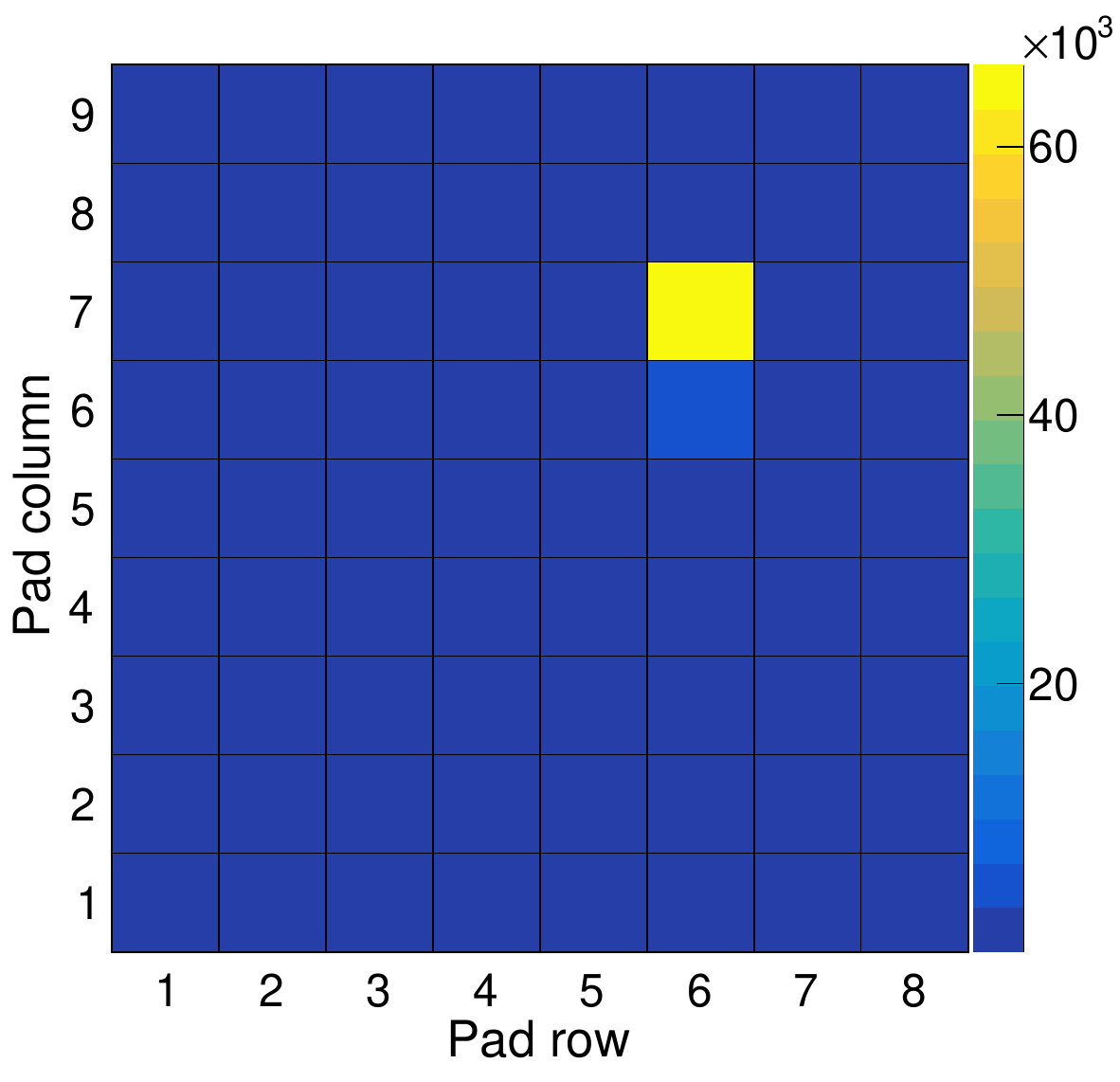}
\qquad
\includegraphics[width=0.43\textwidth]{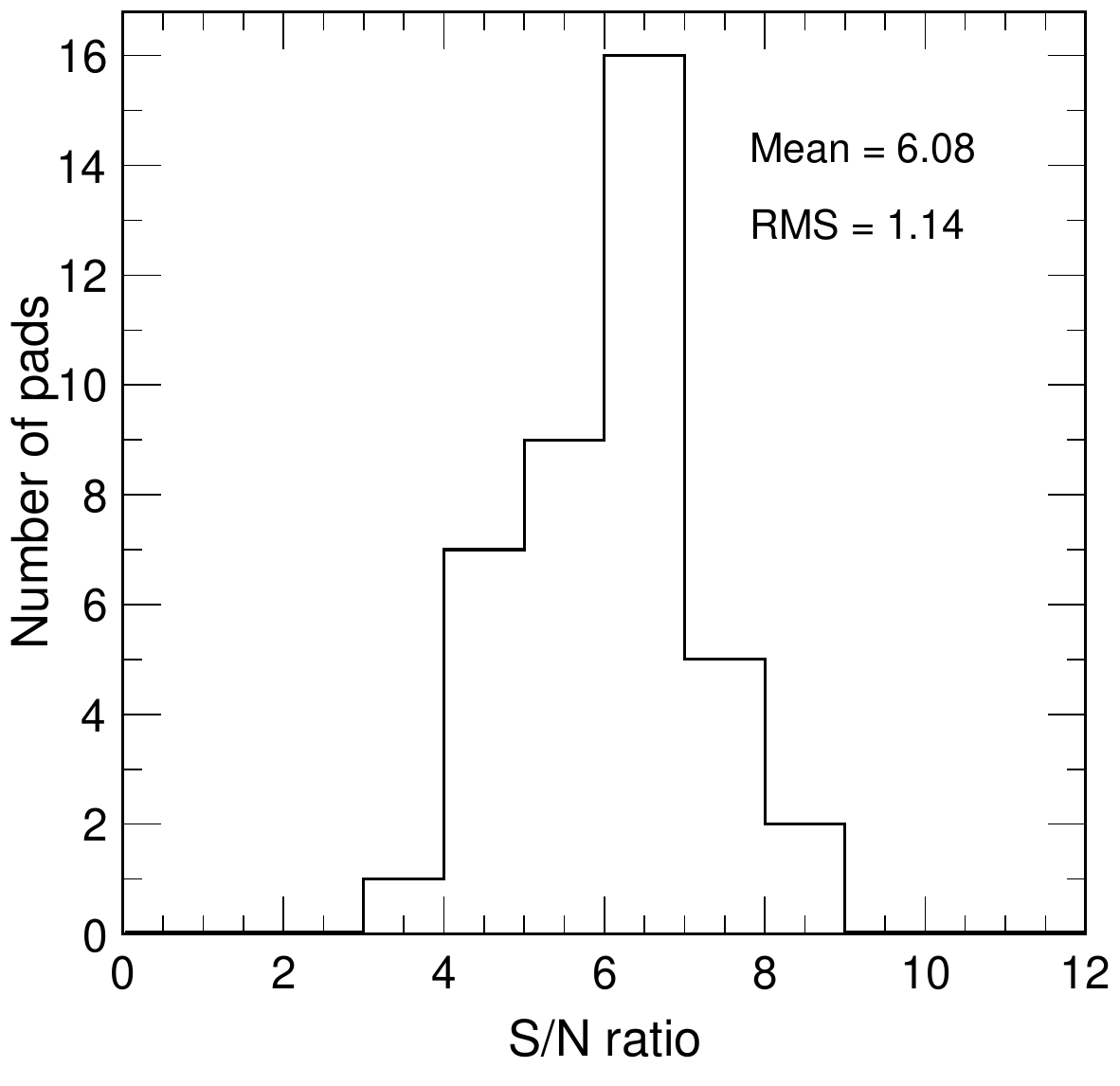}
\caption{(Left) Hit map for a 10~GeV pion beam focused on a single pad of the 8~$\times$~9 silicon pad array. (Right) Signal-to-noise (S/N) ratio across different pads in the detector.\label{fig:position_scan_beam}}
\end{figure}

Similar to the laboratory tests with the $^{90}$Sr source, a position scan was performed using a 10~GeV pion beam. Since the beam position was fixed, the detectors were moved in both the X and Y directions using a motorized table, which also supported the scintillators. A hit map of the pion beam incident on one of the pads is shown in the left panel of Fig.~\ref{fig:position_scan_beam}, where the color scale represents the number of pion hits, scaled by the deposited energy in ADC values. The scan was conducted at a fixed reverse bias voltage of 220~V. The signal-to-noise (S/N) ratios observed for different detector pads are shown in the right panel of Fig.~\ref{fig:position_scan_beam}. The average S/N ratio across the pads is approximately 6.1. These results demonstrate a uniform response across the detector pads.

\subsection{Detector response to positron beam}
When a charged particle travels through an absorber material, it loses its energy through various electromagnetic interactions. The amount of energy loss depends on the particle's mass, velocity, and the material's properties. For high-energy particles, radiative losses dominate over the collisional and ionization losses. In a tungsten absorber of 3.5 mm thickness (1 radiation length, X$_0$), the critical energy threshold for radiative losses is 8 MeV\cite{LeoBook}. Given that the positron beam energy (2–4 GeV) far exceeds this threshold, its primary energy loss mechanisms are bremsstrahlung and pair production, which produce a cascading effect, forming an electromagnetic shower. This shower was studied using a single detector layer while varying the number of absorber layers to probe different depths.

\subsubsection{Pedestal calculation}
\begin{figure}[!htbp]
\centering
\includegraphics[width=0.44\textwidth]{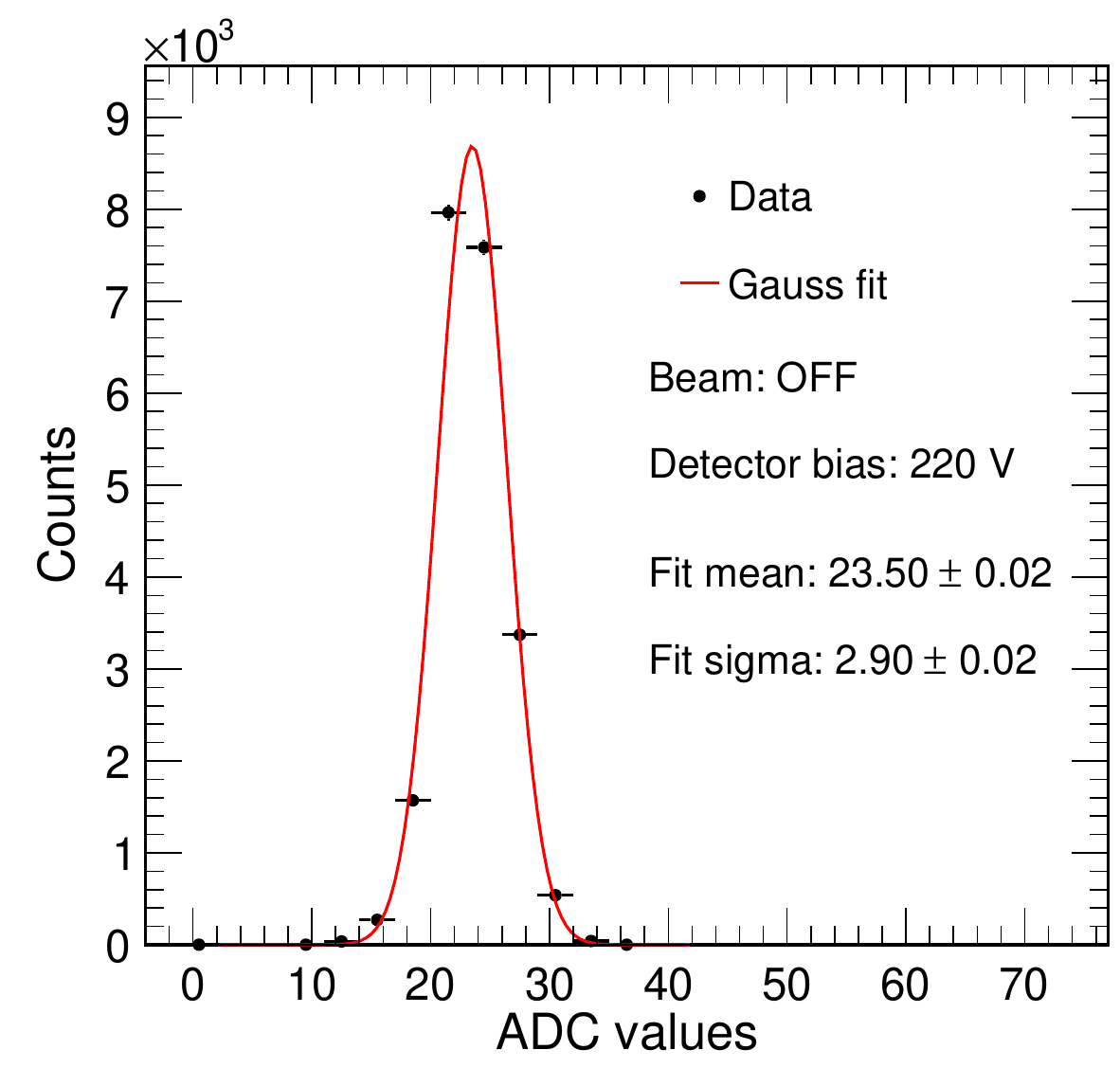}
\qquad
\includegraphics[width=0.44\textwidth]{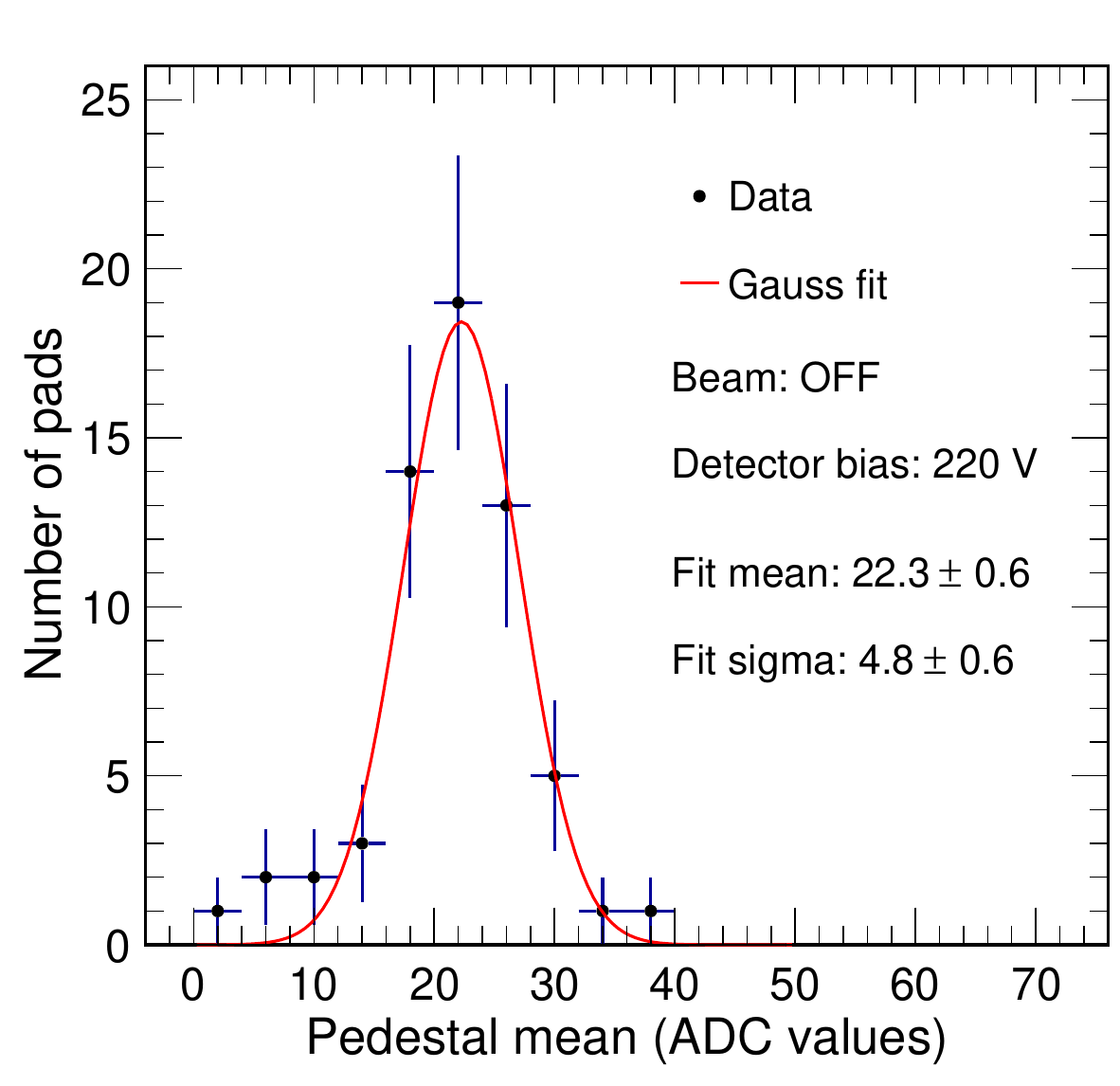}
\caption{(Left) Pedestal distribution for a single detector pad, fitted with a Gaussian function. (Right) Mean pedestal values for all working pads in the detector. The error bars in both figures represent statistical uncertainties.\label{fig:pedestal_analysis}}
\end{figure}

To estimate the pedestal, data were collected under beam-off conditions at a reverse bias voltage of 220~V. The left panel of Fig.~\ref{fig:pedestal_analysis} shows the pedestal distribution for a single detector pad, which follows a Gaussian shape and is therefore fitted with a Gaussian function. The mean values obtained from such fits are shown for all 61 functional detector pads in the right panel of Fig.~\ref{fig:pedestal_analysis}. This excludes the 7 pads with non-functional HGCROC channels and the 4 additional pads with no response. This distribution of mean pedestal values is also fitted with a Gaussian function, yielding an average pedestal value of approximately 22.3 ADC values. For subsequent analysis, the mean pedestal and three times its standard deviation, along with the common-mode noise~\cite{focal_ptype_published}, are subtracted on a pad-by-pad basis for each event to effectively remove the noise contribution.

\subsubsection{Energy deposition and calibration}
The energy deposited by the electromagnetic shower (measured in ADC values) in each radiation length is calculated using a clustering algorithm in both data and simulations. For each event, the pad with the highest energy deposition after pedestal subtraction is identified as the seed. Energy is then summed over all connected neighboring pads to suppress uncorrelated or isolated signals. To account for split clusters, if multiple clusters are found within a 6x6 pad region centered around the beam axis, the clusters were merged provided that the distance of the split cluster from the main cluster is within 4~cm. This clustering algorithm yields two key observables: the cluster energy and the cluster size. The cluster energy gives the energy deposited by the electromagnetic shower, and the cluster size reflects the transverse spread of the shower perpendicular to the beam direction.

\begin{figure}[!htbp]
\centering
\includegraphics[width=0.45\textwidth]{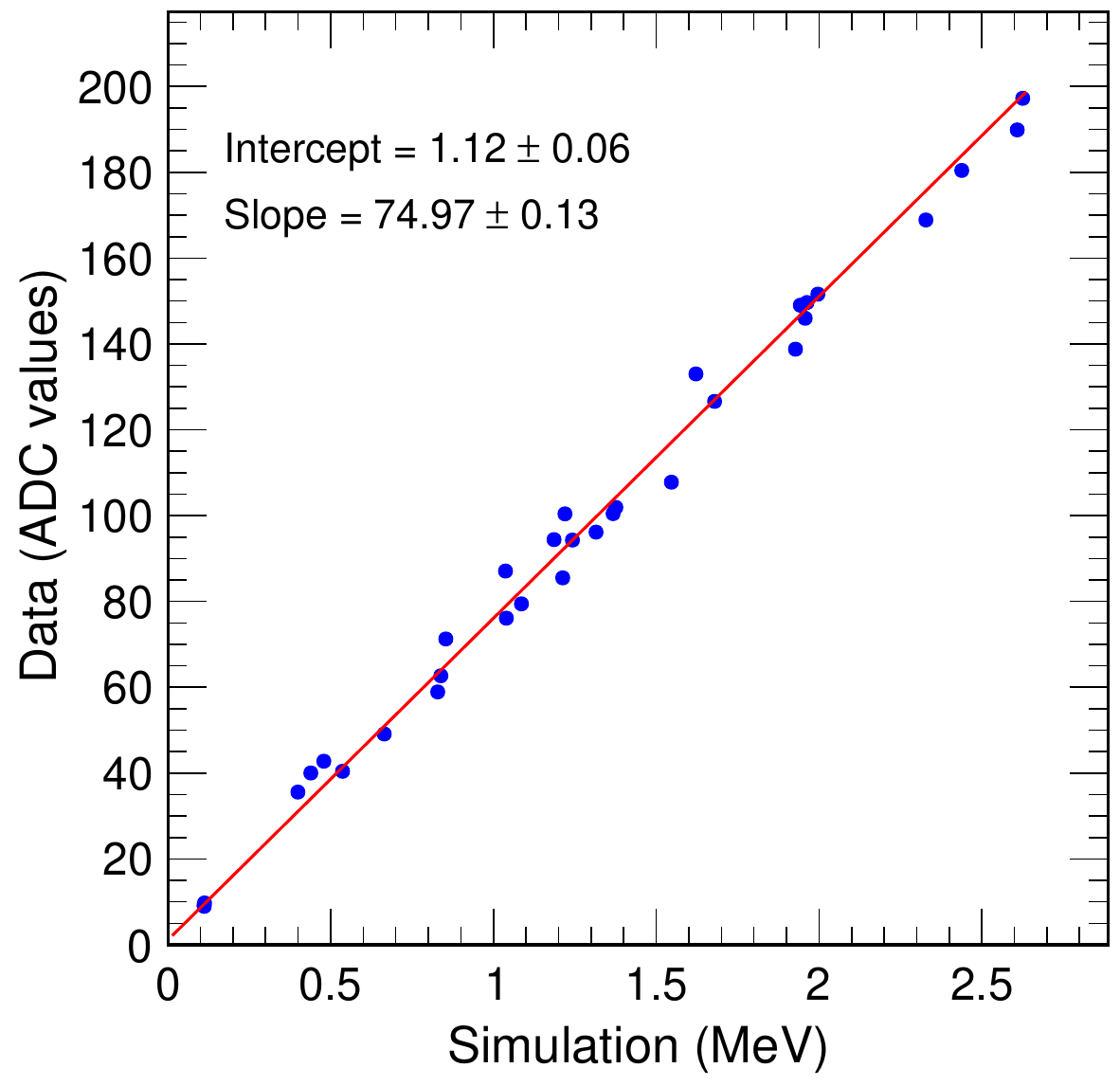}
\caption{The energy deposited by electrons in the detector in experimental data (ADC values) against the deposited energy in simulations (MeV) for 2-4 GeV $\mathrm{e^+}$ beam at 0-10 X$_0$. A linear trend observed and fitted with first-order polynomial. Due to large statistics, the error bars representing statistical uncertainties are small and within the marker size.\label{fig:slope}}
\end{figure}

Since the measured energy is in ADC units, a conversion to physical energy units (MeV) is necessary for comparison with simulation results. This is achieved by comparing the mean deposited energy per radiation length for 2-4 GeV $\mathrm{e^+}$ beam with that obtained from simulations, as shown in Fig.~\ref{fig:slope}. The resulting distribution is fitted with a first-order polynomial. The slope and intercept obtained from the fit are then used to convert the deposited energy from ADC values to MeV in the experimental data.

\subsubsection{Longitudinal shower profile}
\begin{figure}[htp]
\centering
\includegraphics[width=0.47\textwidth]{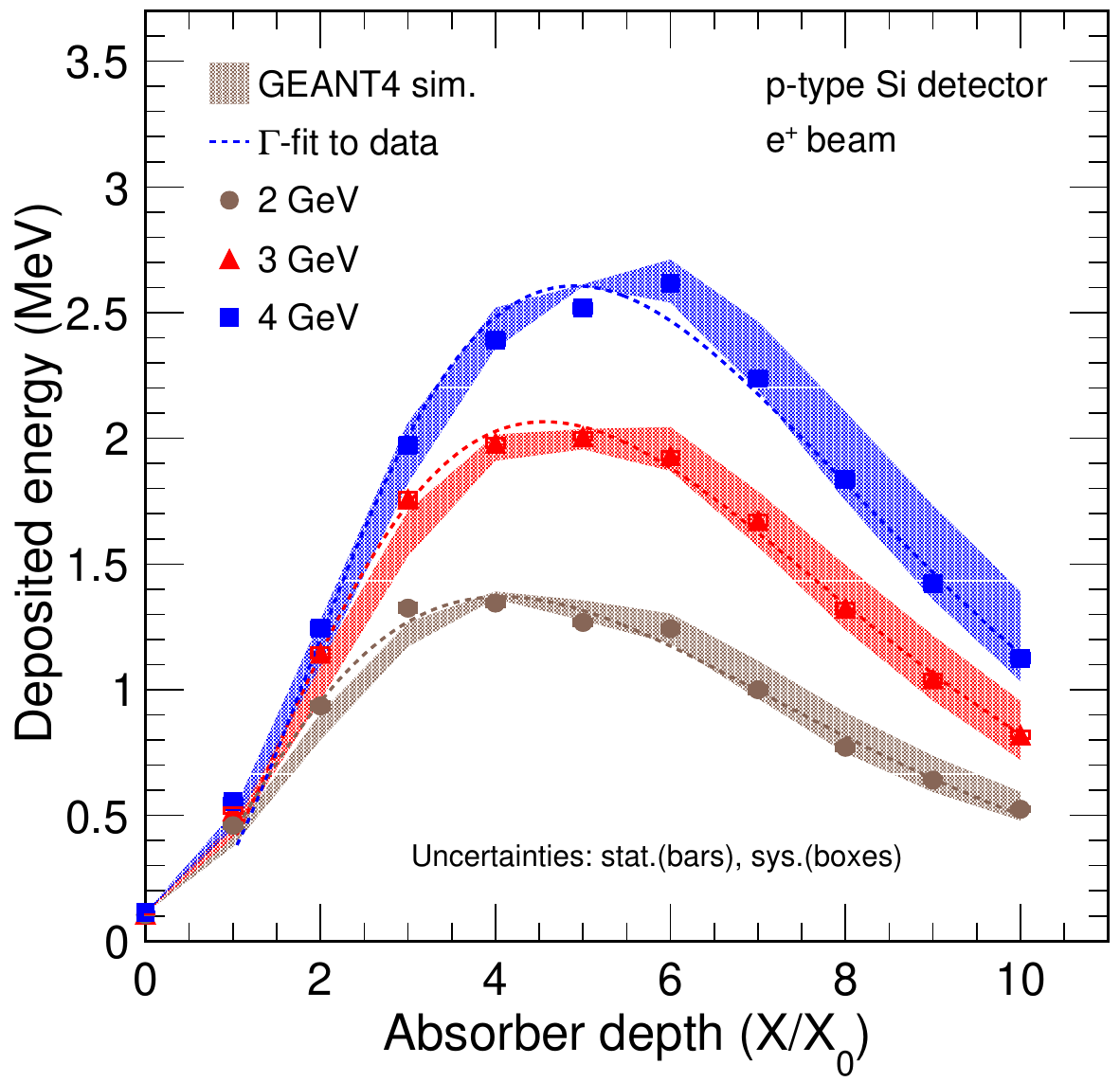}
\qquad
\includegraphics[width=0.47\textwidth]{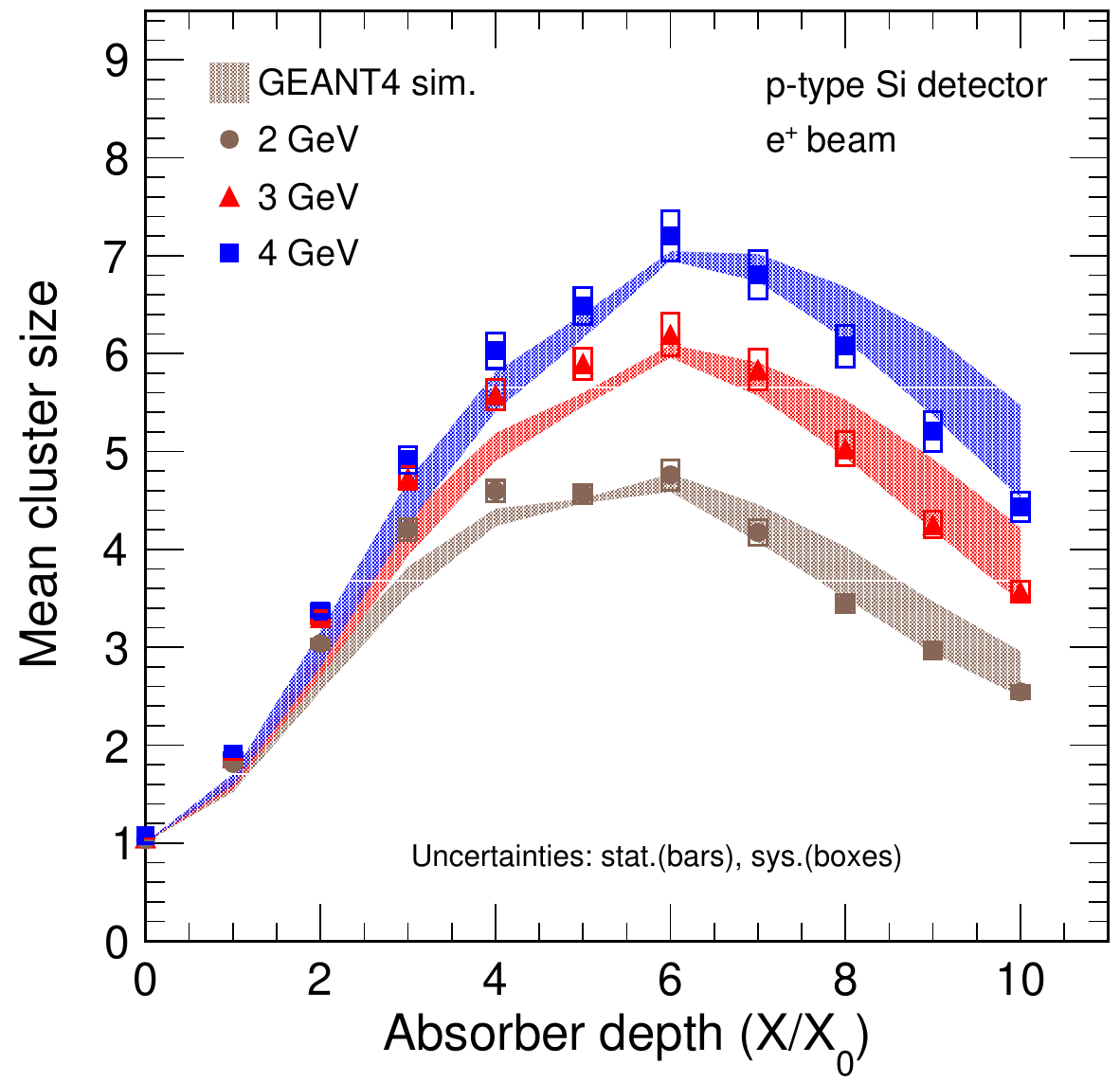}
\caption{Longitudinal shower profile (left) and mean cluster size distribution (right) for electrons with energies ranging from 2~GeV to 4~GeV. Markers represent experimental data, while shaded bands indicate simulation results. The longitudinal shower profiles are fitted using a $\Gamma$-distribution, shown with dotted lines. Error bars representing statistical uncertainties are within the marker size; systematic uncertainties are shown as boxes for data and as bands for simulations.\label{fig:shower_profile}}
\end{figure}

After the energy calibration, the mean energy deposited by the electromagnetic shower is plotted as a function of absorber depth, referred to as longitudinal shower profile, as shown in the left panel of Fig.~\ref{fig:shower_profile}. The absorber depth is expressed in units of radiation length (X$_0$), where one 3.5 mm thick tungsten plate corresponds to 1 X$_0$. The shower profile indicates that the mean deposited energy increases with depth as the electromagnetic shower develops, reaches a peak known as the shower maximum, and then decreases as the photon energies fall below the pair production threshold and ionization losses dominate over bremsstrahlung for electrons. The data is fitted with a Gamma distribution\cite{LeoBook}, which yields shower maximum positions at 3.9 X$_0$, 4.2 X$_0$, and 4.6 X$_0$ for beam energies 2-4 GeV. This shift of the shower maximum to larger depths reflects the deeper penetration of higher-energy showers. The measured shower profiles are compared with Geant4 simulations, which show good agreement with the data.

Systematic uncertainties in the data are estimated by comparing two analysis methods: summing all clusters within a 6x6 region near the beam axis (to account for split clusters) versus analyzing events using only single clusters. In simulations, systematic uncertainty is evaluated by varying the absorber depth by 5\%, accounting for any unaccounted upstream material. These uncertainties are represented by open boxes in the data and shaded bands in the simulation.

To estimate the transverse size of the shower, the mean cluster size is plotted as a function of radiation length, as shown in the right panel of Fig.~\ref{fig:shower_profile}. The shower size increases with radiation length due to multiple scattering between shower particles, the finite opening angle between electron-positron pairs produced during pair production, and the emission of bremsstrahlung photons at different angles. These effects collectively contribute to the lateral spread of the shower. The transverse size reaches a maximum and then decreases as the energies of secondary particles fall below the critical threshold, leading to their absorption. The measured cluster sizes are compared with simulations, which show good agreement with the data.

\section{Summary}
The p-type silicon pad array detectors were successfully fabricated for the first time in India at Semi-Conductor Laboratory (SCL), Mohali. 
Electrical characterization (IV and CV measurements) done on the pad array shows that the leakage current is below 50~nA/pad up to 450~V, the pad capacitance is below 40~pF, the full depletion voltage is under 150~V, and the breakdown voltage exceeds 450~V for most of the pads in the array.

Following packaging, the detectors were characterized in the laboratory using a $^{90}$Sr $\beta^-$ source, where a clear MIP signal with a signal-to-noise ratio above 5.5 was observed. Voltage and position scans were also conducted, which helped in the determination of the optimal operating voltage and confirmed the uniform response across different detector pads.

After satisfactory laboratory measurements, beam tests were conducted at the CERN PS using 10~GeV pion and 2–4~GeV positron beams. A distinct minimum ionizing particle (MIP) signal was observed with the pion beam. Voltage scans showed a saturation in the MPV-to-pedestal separation around FDV consistent with laboratory results, while position scans yielded an average signal-to-noise (S/N) ratio of approximately 6.1. For performance evaluation, the detectors were exposed to a positron beam, allowing the study of electromagnetic shower energy and size. By varying the number of absorber plates placed in front of a single detector layer, the shower was sampled at different depths, generating the longitudinal shower profiles. The measured shower energy and size were compared with Geant4 simulations, showing good agreement with each other. Overall, the detector exhibited stable and uniform performance in both laboratory and beam test environments, demonstrating its suitability for future applications such as the FoCal experiment in ALICE.

Building upon the promising results, further TCAD simulations and process optimizations are being done with three guard rings to improve the S/N ratio and achieve higher breakdown voltage over all 72 detector pads. It is expected to observe improvement in the next iteration of detector fabrication. Additionally, later this year, a full-scale prototype comprising 20 Si-W detector layers will be developed to assess the energy resolution of the sampling calorimeter. This prototype will be tested with pion and electron beams at the Super Proton Synchrotron (SPS) at CERN.

\acknowledgments
The authors would like to thank the Semi-Conductor Laboratory (SCL), Mohali team for the design, fabrication and packaging of the silicon pad array detector; Micropack and KHMDL, Bangalore, for PCB fabrication and component assembly. We thank Tatsuya~Chujo, Ian~Gardner~Bearden, Nicola~Minafra, Tommaso~Isidori, Ton~Van~Den~Brink, the CERN PS accelerator team, LPSC Grenoble team and the ALICE FoCal collaboration members for their constant support throughout the project work and during the test beam. We also thank DAE and DST India for their financial support through the project entitled "Indian participation in the ALICE experiment at CERN". The work is also partly funded through the J.C. Bose fellowship of DST, awarded to BM. The authors acknowledge the use of the Garuda and Kanaad HPC facilities at the School of Physical Sciences, NISER.


\bibliographystyle{JHEP}
\bibliography{biblio.bib}

\providecommand{\href}[2]{#2}\begingroup\raggedright\begin{thebibliography}{10}

\bibitem{Si_detector_medicalImaging}
L.M.~Montaño~Zetina, \emph{Silicon detectors applied to medical imaging}, \href{https://doi.org/10.1063/1.1604091}{\emph{AIP Conference Proceedings} {\bfseries 674} (2003) 344}.

\bibitem{Si_detector_application}
H.-W.~Sadrozinski, \emph{Applications of silicon detectors}, \href{https://doi.org/10.1109/23.958703}{\emph{IEEE Transactions on Nuclear Science} {\bfseries 48} (2001) 933}.

\bibitem{Si_detector_highEnergyPhysics}
H.-G.~Moser, \emph{Silicon detector systems in high energy physics}, \href{https://doi.org/https://doi.org/10.1016/j.ppnp.2008.12.002}{\emph{Progress in Particle and Nuclear Physics} {\bfseries 63} (2009) 186}.

\bibitem{focaltdr}
{\scshape ALICE} collaboration, \emph{{Technical Design Report of the ALICE Forward Calorimeter (FoCal)}},  Tech. Rep. \href{https://cds.cern.ch/record/2890281}{CERN-LHCC-2024-004, ALICE-TDR-022}, CERN, Geneva (2024).

\bibitem{Clice_pads1}
{\scshape CALICE} collaboration, \emph{{Design and Electronics Commissioning of the Physics Prototype of a Si-W Electromagnetic Calorimeter for the International Linear Collider}}, \href{https://doi.org/10.1088/1748-0221/3/08/P08001}{\emph{JINST} {\bfseries 3} (2008) P08001} [\href{https://arxiv.org/abs/0805.4833}{{\ttfamily 0805.4833}}].

\bibitem{Clice_pads2}
{\scshape CALICE} collaboration, \emph{{Response of the CALICE Si-W Electromagnetic Calorimeter Physics Prototype to Electrons}}, \href{https://doi.org/10.1088/1742-6596/160/1/012065}{\emph{J. Phys. Conf. Ser.} {\bfseries 160} (2009) 012065}.

\bibitem{focal_ptype_published}
M.~Aehle, J.~Alme, C.~Arata, I.~Arsene, I.~Bearden, T.~Bodova et~al., \emph{Performance of the electromagnetic and hadronic prototype segments of the alice forward calorimeter}, \href{https://doi.org/10.1088/1748-0221/19/07/P07006}{\emph{Journal of Instrumentation} {\bfseries 19} (2024) P07006}.

\bibitem{pixel_paper1}
A.P.~de~Haas et~al., \emph{{The FoCal prototype\textemdash{}an extremely fine-grained electromagnetic calorimeter using CMOS pixel sensors}}, \href{https://doi.org/10.1088/1748-0221/13/01/P01014}{\emph{JINST} {\bfseries 13} (2018) P01014} [\href{https://arxiv.org/abs/1708.05164}{{\ttfamily 1708.05164}}].

\bibitem{Pixel_paper2_calice}
K.~Kawagoe et~al., \emph{{Beam test performance of the highly granular SiW-ECAL technological prototype for the ILC}}, \href{https://doi.org/10.1016/j.nima.2019.162969}{\emph{Nucl. Instrum. Meth. A} {\bfseries 950} (2020) 162969} [\href{https://arxiv.org/abs/1902.00110}{{\ttfamily 1902.00110}}].

\bibitem{sanjib_paper}
S.~Muhuri et~al., \emph{{Fabrication and beam test of a silicon-tungsten electromagnetic calorimeter}}, \href{https://doi.org/10.1088/1748-0221/15/03/P03015}{\emph{JINST} {\bfseries 15} (2020) P03015} [\href{https://arxiv.org/abs/1911.00743}{{\ttfamily 1911.00743}}].

\bibitem{focal_LOI}
{\scshape ALICE} collaboration, \emph{{Letter of Intent: A Forward Calorimeter (FoCal) in the ALICE experiment}},  Tech. Rep. \href{https://cds.cern.ch/record/2719928}{CERN-LHCC-2020-009, LHCC-I-036}, CERN, Geneva (2020).

\bibitem{physics_of_focal}
{\scshape ALICE} collaboration, \emph{{Physics performance of the ALICE Forward Calorimeter upgrade}},  Tech. Rep. \href{https://cds.cern.ch/record/2869141}{ALICE-PUBLIC-2023-004}, CERN, Geneva (2023).

\bibitem{HGCROCv2_paper}
{\scshape CMS} collaboration, \emph{{Performance study of HGCROC-v2: the front-end electronics for the CMS High Granularity Calorimeter}}, \href{https://doi.org/10.1088/1748-0221/15/04/C04055}{\emph{JINST} {\bfseries 15} (2020) C04055}.

\bibitem{ALPIDE_chip}
G.~{Aglieri Rinella}, \emph{The alpide pixel sensor chip for the upgrade of the alice inner tracking system}, \href{https://doi.org/https://doi.org/10.1016/j.nima.2016.05.016}{\emph{Nuclear Instruments and Methods in Physics Research Section A: Accelerators, Spectrometers, Detectors and Associated Equipment} {\bfseries 845} (2017) 583}.

\bibitem{ntype_fabrication_paper}
Sawan, G.~Tambave, J.~Bouly, O.~Bourrion, T.~Chujo, A.~Das et~al., \emph{Design, fabrication and characterization of 8x9 n-type silicon pad array for sampling calorimetry}, \href{https://doi.org/10.1088/1748-0221/20/05/P05007}{\emph{Journal of Instrumentation} {\bfseries 20} (2025) P05007}.

\bibitem{ntype_testbeam_niser}
Sawan, J.~Bouly, O.~Bourrion, M.~Bregant, A.~van~den Brink, T.~Chujo et~al., \emph{Beam test of n-type silicon pad array detector at ps cern}, \href{https://doi.org/10.1088/1748-0221/19/09/P09016}{\emph{Journal of Instrumentation} {\bfseries 19} (2024) P09016}.

\bibitem{TCAD_simulations}
E.~Guichard and I.~Silvaco, \emph{Silvaco tcad},  Sep, 2022.
\newblock doi:10.21981/MZFR-HK34.

\bibitem{scl}
``{Semi-Conductor Laboratory, India}.'' \url{https://www.scl.gov.in/}.

\bibitem{bourrion2023prototype}
O.~Bourrion, D.~Tourres, R.~Guernane, C.~Arata, J.L.~Bouly and N.~Ponchant, \emph{{Prototype electronics for the silicon pad layers of the future Forward Calorimeter (FoCal) of the ALICE experiment at the LHC}}, \href{https://doi.org/10.1088/1748-0221/18/04/P04031}{\emph{JINST} {\bfseries 18} (2023) P04031} [\href{https://arxiv.org/abs/2302.13912}{{\ttfamily 2302.13912}}].

\bibitem{KCU105}
``{{AMD Kintex UltraScale FPGA KCU105} Evaluation Kit}.'' \url{https://www.xilinx.com/products/boards-and-kits/kcu105.html}.

\bibitem{lower_breakdown2}
J.C.~Gallagher, M.A.~Mastro, A.G.~Jacobs, R.J.~Kaplar, K.D.~Hobart and T.J.~Anderson, \emph{Detecting defects that reduce breakdown voltage using machine learning and optical profilometry}, \href{https://doi.org/10.1038/s41598-024-57875-5}{\emph{Scientific Reports} {\bfseries 14} (2024) 7440}.

\bibitem{T10_beam}
G.~D’Alessandro, S.T.~Boogert, S.~Gibson, L.J.~Nevay, W.~Shields, J.~Bernhard et~al., \emph{Implementation of cern secondary beam lines t9 and t10 in bdsim}, \href{https://doi.org/10.1088/1742-6596/1350/1/012095}{\emph{Journal of Physics: Conference Series} {\bfseries 1350} (2019) 012095}.

\bibitem{mixed-target}
``{Mixed-Target}.'' \url{http://sba.web.cern.ch/sba/targets/TargetNorth.html}.

\bibitem{geant4package}
{\scshape GEANT4} collaboration, \emph{{GEANT4--a simulation toolkit}}, \href{https://doi.org/10.1016/S0168-9002(03)01368-8}{\emph{Nucl. Instrum. Meth. A} {\bfseries 506} (2003) 250}.

\bibitem{LeoBook}
W.R.~Leo, \emph{{{Techniques for nuclear and particle physics experiments: a how-to approach}}}, Springer, Berlin, 2~ed. (1994), \href{https://doi.org/10.1007/978-3-642-57920-2}{10.1007/978-3-642-57920-2}.

\end{thebibliography}\endgroup
\end{document}